\documentclass[draftclsnofoot, 12pt,onecolumn,oneside,compress]{IEEEtran}

\ifCLASSINFOpdf
\else
   \usepackage[dvips]{graphicx}
\fi
\usepackage{url}

\usepackage{graphicx,balance}
\usepackage{bm}
\usepackage{amssymb}
\usepackage{amsmath}
\usepackage{textcomp}
\usepackage{color}
\usepackage{epstopdf}
\usepackage{subfigure}
\usepackage{booktabs}
\usepackage{enumitem}
\usepackage{algorithm}
\usepackage{algorithmic}
\usepackage{bbm}
\usepackage{makecell}
\usepackage{multirow}

\begin{document}

\title{Computer Vision Aided mmWave Beam Alignment in V2X Communications}

\author{Weihua Xu, Feifei Gao, Xiaoming Tao, Jianhua Zhang, and Ahmed Alkhateeb
\thanks{W. Xu and F. Gao are with Institute for Artificial Intelligence, Tsinghua
University (THUAI), Beijing National Research Center for Information
Science and Technology (BNRist), Department of Automation, Tsinghua
University, Beijing, P.R. China, 100084 (email: xwh19@mails.tsinghua.edu.cn,
feifeigao@ieee.org).}
\thanks{X. Tao is with the Department of Electronic Engineering, Tsinghua University, Beijing, P.R. China, 100084 (email: taoxm@tsinghua.edu.cn).}
\thanks{J. Zhang is with the State Key Laboratory of Networking and Switching Technology, Beijing University of Posts
and Telecommunications, Beijing 100876, China (e-mail: jhzhang@bupt.edu.cn).}
\thanks{A. Alkhateeb is with the Department of Electrical and Computer Engineering, University of Texas at Austin, Austin, TX 78712-1687 USA (e-mail:
alkhateeb@asu.edu).}
}

\maketitle
\vspace{-5mm}

\begin{abstract}
Visual information, captured for example by cameras, can effectively reflect the sizes and locations of the environmental scattering objects, and thereby can be used to infer communications parameters like propagation directions, receiver powers, as well as the blockage status. In this paper, we propose a novel beam alignment framework that leverages images taken by cameras installed at the mobile user. Specifically, we utilize 3D object detection techniques to extract the size and location information of the dynamic vehicles around the mobile user, and design a deep neural network (DNN) to infer the optimal beam pair for transceivers without any pilot signal overhead. Moreover, to avoid performing beam alignment too frequently or too slowly, a beam coherence time (BCT) prediction method is developed based on the vision information. This can effectively improve the transmission rate compared with the beam alignment approach with the fixed BCT. Simulation results show that the proposed vision based beam alignment methods outperform the existing LIDAR and vision based solutions, and demand for much lower hardware cost and communication overhead.
\end{abstract}


\IEEEpeerreviewmaketitle

\newpage

\section{Introduction}
Beamforming has been deemed as the critical technique to  overcome the signal attenuation of high frequency band, especially for the millimeter  wave or even Terahertz communication \cite{Roh}. Traditional beamforming strategies include sweeping the beam with a certain codebook \cite{Noh} to maximize the receive signal-to-noise ratio (SNR) or calculating the beamforming matrix directly from the estimated channel matrix \cite{Sohrabi}. However, it is generally known that producing high gain beam with large antenna array leads to the huge time and spectrum overhead. Recently, integrated sensing and communication (ISAC) has drawn great attention for its capability to assist beamforming \cite{Heath}-\cite{Xu2}. The principle behind is that the sensor data, i.e., the point cloud or the RGB/depth images from GPS, Radar, LIDAR, and camera equipped at many intelligent terminals, are the valid out-of-band  information that can indicate the spatial characteristics of the communications environment.

In \cite{Wang}, the authors proposed to use the locations of the mobile station (MS) and the surrounding vehicles to realize power estimation of all the beam pairs under the V2X scenario. Although the high throughput ratio can be achieved by \cite{Wang}, the requirement for all surrounding vehicles to transmit the instantaneous locations to the MS will introduce the huge cost and delay. Thus, the \emph{proactive} perception capability of the BS is further explored to reduce the transmission overhead of surrounding vehicles as well as to obtain more accurate and abundant environmental information. With the aid of the Radar mounted at BS, \cite{Gonz} proposes a new hybrid beamforming scheme, and \cite{FLiu} presents a beam tracking method in multi-user situation. The authors in \cite{UDemirhan} propose a Radar-aided beam alignment method, and verify it by the real-world dataset. Nevertheless, the detection accuracy of Radar is inferior to the LIDAR and camera, as the Radar signal is mainly suitable to detect the directional information of transceivers but cannot accurately reflect the precise distribution of the surrounding scatters. In \cite{Dias}-\cite{klautau}, the Base Station (BS) and MS use LIDAR to scan point clouds for accelerating the beam alignment. In \cite{Mashhadi}, the authors proposed a federated learning based beam alignment method to use the LIDAR data from multiple distributed MSs and train a DNN cooperatively for realizing better beam alignment accuracy. However, LIDAR is expensive and is not universally implemented compared to the ordinary cameras.

The authors in \cite{Ahmed1} then proposed to equip the BS with multiple cameras and take the RGB images of the user and surrounding vehicles. The beam sequence and the image coordinates of environmental objects are utilized as the input feature to predict blockage and BS handover from the recurrent neural network (RNN). In \cite{YTian}-\cite{SJiang}, the previous images taken at BS and the beam sequences are utilized for beam tracking. In \cite{GCharan}, a multi-modal beam alignment method is proposed by the conjunctive use of the images taken at BS and the location coordinates of MS.

Nevertheless, if BS is obliged to take images, then additional characteristic information of the MS is required by BS to identify the MS from all detected objects inside one image, i.e., to match the MS's radio signal with the corresponding visual information \cite{Pinho}. Therefore, in \cite{Ahmed1}-\cite{GCharan}, the BS needs to rely on the previous beam sequences or the location fed back by MS to achieve the blockage/handover prediction or beam tracking/selection, which leads to inevitable communication overhead. When the MS is blocked by the surrounding objects in the camera view of BS, the visual information of the MS will be lost to cause the degradation of the beamforming performance. Moreover, taking image at BS may violate the privacy of the customers and may not be accepted in real implementation.

Thus, an alternative way is to utilize the vision capability of the intelligent MS, such as autonomous car and unmanned aerial vehicle, to avoid the overhead for MS identification and obtain better beamforming performance. In fact, the visual resources at MS are more convenient to be obtained, especially with the gradual popularization of automatic driving, and UAV reconnaissance, since the visual data are also widely used for navigation and obstacle avoidance. In \cite{Xu2}, the authors use the scene images taken at MS to infer the channel covariance matrix. However, the covariance matrix is only the statistical characteristic of channel, whereas the instantaneous beam alignment relying on the vision of user terminal has not been tackled yet, to the best of the authors' knowledge.

In this paper, we propose to utilize the camera images taken at MS for the beam alignment under dynamic environment. Specifically, the main contributions of this paper contains following three different aspects:
\begin{itemize}
    \item[1)]
    Vision based beam alignment when the MS location is available (VBALA), in which the 3D detection technique is utilized to obtain the 3D spatial distribution information of MS surrounding dynamic objects, and then a DNN is designed to predict the optimal beam pair from the MS location and vehicle distribution information. Vision based beam alignment when the MS location is unavailable or not accurate enough due to the strong noise corruption (VBALU), in which the optimal beam is predicted only through the images.
    \item[2)]
    Moreover, we present a vision based method to predict the beam coherent time (BCT) (VPBCT), i.e., the duration that the optimal beam pair remains unchanged for codebook-based beamforming.
    \item[3)]
    By the image and channel data generated from the autonomous driving and ray tracing simulation software, the proposed VBALA and VBALU are verified to achieve better beam alignment performance than the Lidar and BS's vision based methods, which demonstrates the advantage of utilizing the images taken at MS. The proposed VPBCT can outperform the fixed BCT based beamforming approach. The simulated dataset is made available to the public \cite{github}.
\end{itemize}

\begin{table}[t]
\centering
\caption{The related research works for comparison}
\begin{tabular}{|c|c|c|c|c|}
\hline
Paper             & Sensor                                                                 & Sensor Data Type                    & Task                                                                                     & \begin{tabular}[c]{@{}c@{}}Limitations compared with\\ the proposed methods\end{tabular}                              \\ \hline
\cite{Wang}           & \begin{tabular}[c]{@{}c@{}}GPS at\\ MS and other vehicles\end{tabular} & Location coordinates                & Beam power estimation                                                                    & \begin{tabular}[c]{@{}c@{}}Huge cost and delay for transmission\\ of instantaneous locations\end{tabular}             \\ \hline
\cite{Gonz},\cite{UDemirhan}   & \multirow{2}{*}{Radar at BS}                                           & \multirow{2}{*}{Radar echo signals} & Beam alignment                                                                           & \multirow{2}{*}{\begin{tabular}[c]{@{}c@{}}Inferior detection accuracy\\ than the camera\end{tabular}}      \\ \cline{1-1} \cline{4-4}
\cite{FLiu}           &                                                                        &                                     & Beam tracking                                                                            &                                                                                                                       \\ \hline
\cite{Dias}          & LIDAR at BS or MS                                                      & \multirow{2}{*}{Point Clouds}       & \multirow{2}{*}{Beam alignment}                                                          & \multirow{2}{*}{\begin{tabular}[c]{@{}c@{}}More expensive and less universal\\ than the ordinary camera\end{tabular}} \\ \cline{1-2}
\cite{klautau}-\cite{Mashhadi} & LIDAR at MS                                                            &                                     &                                                                                          &                                                                                                                       \\ \hline
\cite{Ahmed1}          & \multirow{3}{*}{Camera at BS}                                          & \multirow{5}{*}{RGB Images}         & \begin{tabular}[c]{@{}c@{}}Blockage prediction\\ and BS handover prediction\end{tabular} & \multirow{3}{*}{\begin{tabular}[c]{@{}c@{}}Requirement for MS identification\\ and privacy concerns\end{tabular}}     \\ \cline{1-1} \cline{4-4}
\cite{YTian}-\cite{SJiang} &                                                                        &                                     & Beam tracking                                                                            &                                                                                                                       \\ \cline{1-1} \cline{4-4}
\cite{GCharan}          &                                                                        &                                     & Beam alignment                                                                           &                                                                                                                       \\ \cline{1-2} \cline{4-5}
\cite{Xu2}          & \multirow{2}{*}{Camera at MS}                                                           &                                     & \multicolumn{1}{l|}{Channel covariance estimation}                                       & \begin{tabular}[c]{@{}c@{}}Only statistical characteristic estimation\\ of channel\end{tabular}                       \\ \cline{1-1} \cline{4-5}
Ours              &                                                                        &                                     & Beam alignment and BCT prediction                                                                          & \multicolumn{1}{l|}{}                                                                                                                      \\ \hline
\end{tabular}
\end{table}

The above related research works are listed in the TABLE~I for clear comparison with the proposed methods.

This paper is organized as follows. Section II introduces the signal model as well as the beam alignment criteria of the communication system. Section III proposes the vision based beam alignment methods with/without MS location, while Section IV presents the vision based BCT prediction method. Section V provides the performance metric, the simulation setup, and numerical results together with detailed discussions. Finally, Section VI draws the conclusions.

Notation: $\bm{A}$ is a matrix; $\bm{\mathcal{A}}$ is a set; $\bm{a}$ is a vector; a is a scalar; $\bm{A}_{[i,j]}$ is the element of the $i$th row and the $j$th column in $\bm{A}$; $\bm{A}_{[i,:]}$ and $\bm{A}_{[:,j]}$ are the $i$th row and the $j$th column of $\bm{A}$ respectively; $\mathcal{N}(\bm{m}_{\mathrm{g}}, \bm{R}_{\mathrm{g}})/\mathcal{CN}(\bm{m}_{\mathrm{g}}, \bm{R}_{\mathrm{g}})$ is the real/complex Gaussian
random distribution with mean $\bm{m}_{\mathrm{g}}$ and covariance $\bm{R}_{\mathrm{g}}$; $\mathrm{Card}(\bm{\mathcal{A}})$ is the cardinality of the set $\bm{\mathcal{A}}$; $\mathrm{E}\{\ \cdot\ \}$ is the expectation operator. For the convenience of expression and reference, the critical notations adopted in the paper are summarized in TABLE~II.

\begin{table}[t]
\centering
\caption{critical notations in the paper}
\begin{tabular}{|c|c|c|c|}
\hline
Notation& Description& Notation& Description\\
\hline
$N_{\mathrm{B}}$ and $N_{\mathrm{U}}$ & Antenna numbers of BS and MS& $\bm{F}$ &  Vehicle distribution feature for VBALA\\
\hline
$K$ & Subcarrier number& $\bm{I}$& Scene image feature for VBALU\\
\hline
$\bm{w}_{\mathrm{B}}$ and $\bm{w}_{\mathrm{U}}$ & Transmit and receive beamforming vector& $\bm{D}$ & Sequence of scene image features for VPBCT\\
\hline
$N_{\mathrm{B}}^{\mathrm{CB}}$ and $N_{\mathrm{U}}^{\mathrm{CB}}$ &  Codebook sizes of transmit and receive beam& $T_{\mathrm{d}}$ & Shooting interval of the camera\\
\hline
$C$ &Number of cameras equipped at MS& $T_{\mathrm{b}}$ & Time for beam alignment during one BCT\\
\hline
$\theta_{\mathrm{M}}^{i}$ & Azimuth of the $i$th camera&  $M$ & Ratio between BCT and shooting interval\\
\hline
$L_{\mathrm{G}}$ and $W_{\mathrm{G}}$ &Length and width of the grid for VBALA& $\bm{\mathcal{Q}}_{\mathrm{T}}$,$\bm{\mathcal{Q}}_{\mathrm{V}}$ and $\bm{\mathcal{Q}}_{\mathrm{E}}$&  \makecell{Index set of image set sequences\\for constructing training, validation and test set}\\
\hline
\end{tabular}
\end{table}

\section{Signal Model}
Let us consider a downlink orthogonal frequency-division multiplexing (OFDM) mmWave communication system with a single BS and a single MS. BS is equipped with a uniform linear array (ULA) of $N_\mathrm{B}$ antennas, and the MS is equipped with a ULA of $N_\mathrm{U}$ antennas. Both the BS and MS are assumed to have only a radio frequency chain. The downlink signal received at the user for the $k$th subcarrier can be expressed as
\begin{equation}
y_{k}=\bm{w}_{\mathrm{U}}^{\mathrm{H}}\bm{H}_k\bm{w}_{\mathrm{B}}s_k+n_k,
\end{equation}
where $s_k\in \mathbb{C}$ is the transmit signal, $\bm{H}_k\in \mathbb{C}^{N_{\mathrm{U}}\times N_{\mathrm{B}}}$ is the downlink channel matrix at the $k$th subcarrier, $\bm{w}_{\mathrm{B}}\in \mathbb{C}^{N_\mathrm{B}\times 1}$ is the transmit beamforming vector, $\bm{w}_{\mathrm{U}}\in \mathbb{C}^{N_{\mathrm{U}}\times 1}$ is the receive beamforming vector, and $n_k \in \mathcal{CN}(0,\sigma^2)$ is the Gaussian noise at the $k$th subcarrier. The transmit signal $s_k$ satisfies $\mathrm{E}\{|s_k|^2\}=P_k$.

From the widely used geometric channel model \cite{AMSa}, the channel matrix $\bm{H}_k$ can be expressed as
\begin{equation}
\bm{H}_k=\sum_{n=0}^{N-1}\sum_{l=1}^{L}\alpha_l e^{-j\frac{2\pi k}{K}n}d(nT_{\mathrm{s}}-\tau_l)\bm{a}_\mathrm{r}(\phi^r_l)\bm{a}_\mathrm{t}^{\mathrm{H}}(\phi^t_l),
\end{equation}
where $\alpha_l$ is the complex gain of the $l$th path, $\tau_l$ is the time delay of the $l$th path, $\phi^\mathrm{r}_l$ and $\phi^\mathrm{t}_l$ are the $l$th path's azimuth angle of arrival and departure respectively, $T_{\mathrm{s}}$ is the sampling interval, $K$ is the number of subcarriers, $N$ is the length of cyclic prefix, and $d(\cdot)$ denotes the pulse shaping filter. Moreover, $\bm{a}_\mathrm{r}(\phi)\in \mathbb{C}^{N_\mathrm{U}\times 1}$ and $\bm{a}_\mathrm{t}(\phi) \in \mathbb{C}^{N_\mathrm{B}\times 1}$ are the complex steering vector of the receive and transmit ULA array respectively.

When the antenna spacing is set as the half carrier wavelength, the mathematical expression of steering vector $\bm{a}_\mathrm{r}(\phi)$ and $\bm{a}_\mathrm{t}(\phi)$ is
\begin{equation}
\begin{aligned}
\bm{a}_\mathrm{r}(\phi)&=\frac{1}{\sqrt{N_{\mathrm{U}}}}[1,e^{j\pi \sin(\phi)},\cdots,e^{j(N_{\mathrm{U}}-1)\pi \sin(\phi)}]^{\mathrm{T}},\\
\bm{a}_\mathrm{t}(\phi)&=\frac{1}{\sqrt{N_{\mathrm{B}}}}[1,e^{j\pi \sin(\phi)},\cdots,e^{j(N_{\mathrm{B}}-1)\pi \sin(\phi)}]^{\mathrm{T}}.
\end{aligned}
\end{equation}
Assume the system utilizes the beamforming codebook to perform analog beamforming. The optimal beam pair $(\bm{w}_{\mathrm{B}}^{\mathrm{opt}},\bm{w}_{\mathrm{U}}^{\mathrm{opt}})$ should be selected from the transmit beam codebook $\bm{\mathcal{W}}_{\mathrm{B}}=\{\bm{w}_{\mathrm{B},1}, \bm{w}_{\mathrm{B},2}, \cdots, \bm{w}_{\mathrm{B},N^{\mathrm{CB}}_{\mathrm{B}}}\}$ and the receive beam codebook $\bm{\mathcal{W}}_{\mathrm{U}}=\{\bm{w}_{\mathrm{U},1}, \bm{w}_{\mathrm{U},2}, \cdots, \bm{w}_{\mathrm{U},N^{\mathrm{CB}}_{\mathrm{U}}}\}$ by maximizing the data rate, i.e.,
\begin{equation}
(\bm{w}_{\mathrm{B}}^{\mathrm{opt}},\bm{w}_{\mathrm{U}}^{\mathrm{opt}})=\mathop{\arg\max}_{(\bm{w}_{\mathrm{B}},\bm{w}_{\mathrm{U}})}\frac{1}{K}\sum_{k=1}^{K}\log_2 \left(1+ \frac{P_k}{\sigma^2}|\bm{w}^{\mathrm{H}}_{\mathrm{U}}\bm{H}_k\bm{w}_{\mathrm{B}}|^2\right).
\end{equation}

\section{Vision Based Beam Alignment}
\begin{figure}[t]
\centering
\includegraphics[width=1\textwidth]{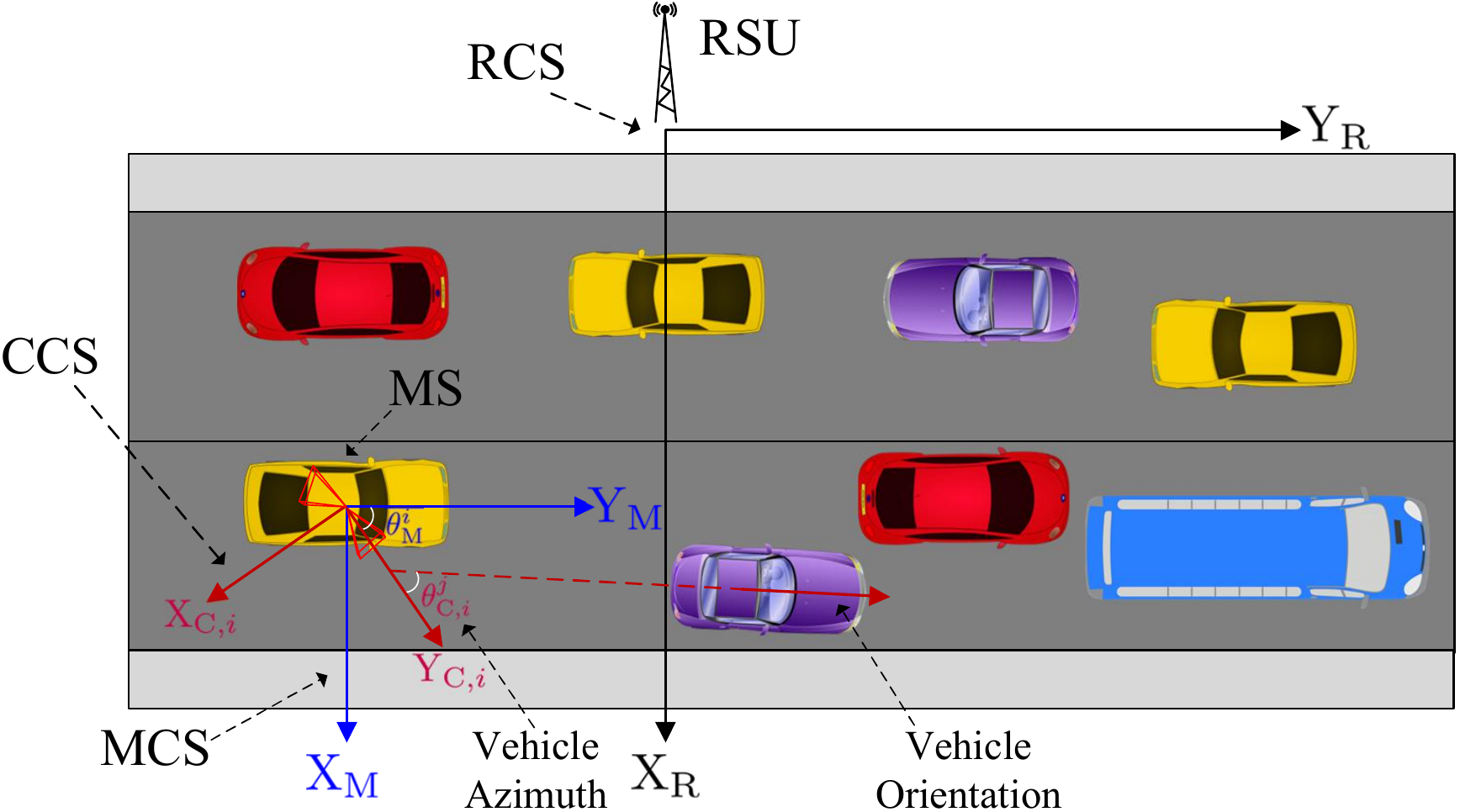}
\caption{The V2X communication scenario for the vision based beam alignment methods.}
\end{figure}

\begin{figure*}[t]
\centering
\includegraphics[width=1\textwidth]{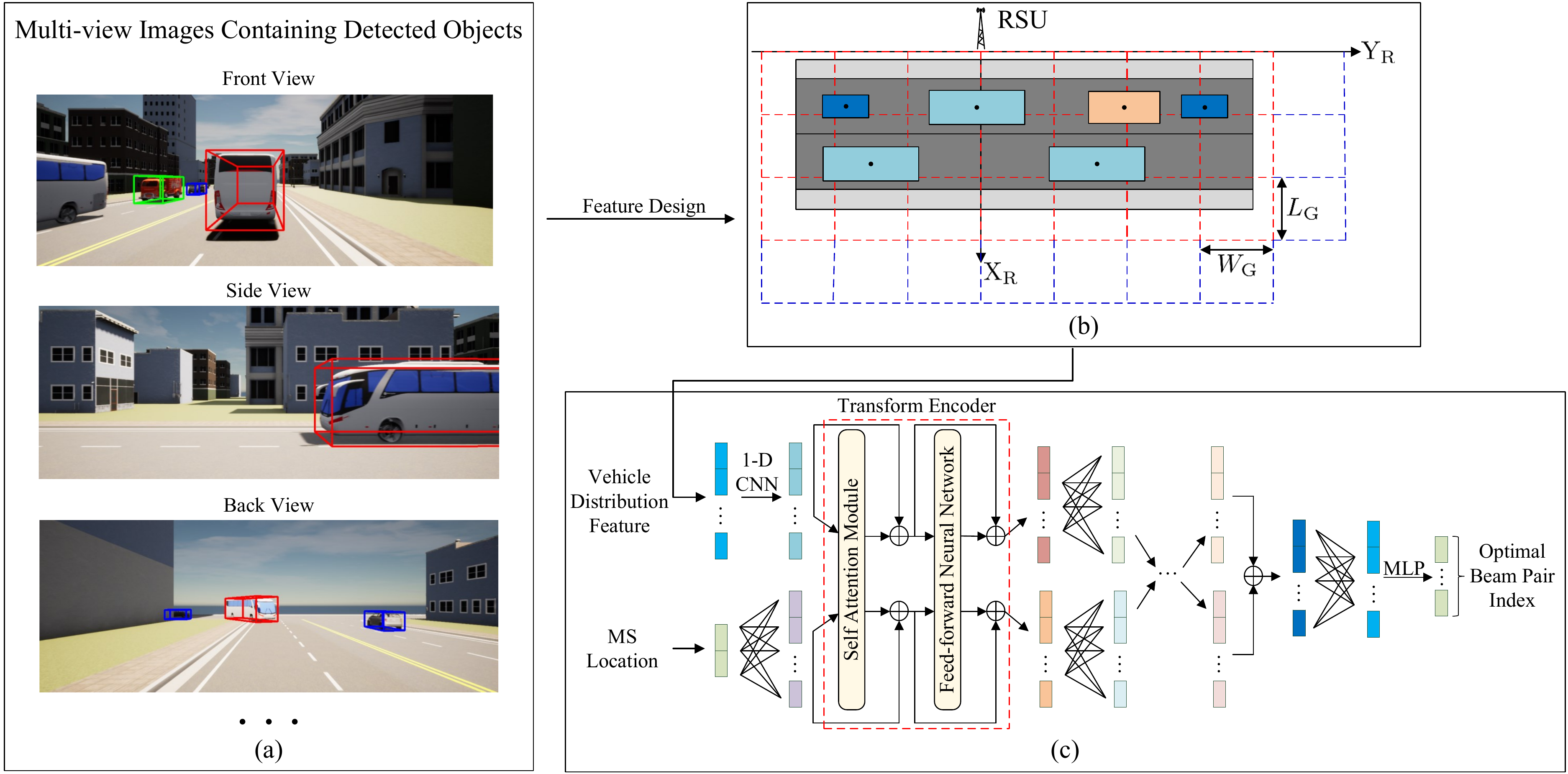}
\caption{The diagram of the proposed VBALA.}
\end{figure*}

The concerned V2X scenario for vision based beam alignment is shown in Fig.~1. The vehicles equipped with many sensors for driver assistance are gradually popularized with the rapid development of autonomous driving technology \cite{MDaily}-\cite{EYurtsever}. We consider the MS is the target vehicle with $C$ auxiliary cameras and runs along the traffic lane, i.e., we assume the MS movement direction is parallel to the traffic lane's right direction. Moreover, there are many other random dynamic vehicles traveling on the same or the nearby lane and acting as the possible blockage objects.

A road-side unit (RSU) acts as the BS and communicates with the MS. The RSUs are expected to be widely deployed in the traffic environments to realize the highly efficient vehicle to infrastructure (V2I) communication for the vehicle safety \cite{AGuerna}, the vehicle network \cite{SBadu} and even edge computing ability \cite{FBusacca}. Here, the key difference from the existing vision based work \cite{Ahmed1}-\cite{GCharan} is that the camera images are taken at MS rather than BS.

The optimal beam alignment between RSU and MS are highly relevant to not only the MS's location but also the size, the location and the number of environmental dynamic scattering objects, e.g., the vehicles around MS. Hence, it is possible for MS to determine the optimal beam pair if the surrounding environment is sufficiently perceived. As Fig.~1 shows, we define an MS coordinate system (MCS) with the $\mathrm{X}_{\mathrm{M}}$-axis, $\mathrm{Y}_{\mathrm{M}}$-axis, and $\mathrm{Z}_{\mathrm{M}}$-axis, in which the MS location, MS moving direction and the traffic lane's vertical direction are the origin, $\mathrm{Y}_{\mathrm{M}}$-axis and $\mathrm{X}_{\mathrm{M}}$-axis respectively, while $\mathrm{Z}_{\mathrm{M}}$-axis is upward and is perpendicular to the traffic lane plane.

All $C$ cameras are ordinary monocular cameras, while the MS could utilize the monocular 3D object detection to extract the sizes, the orientations and the locations of the surrounding vehicles \cite{Detec}. For the $i$th camera, $i=1,2,\cdots,C$, we also define its own camera coordinate system (CCS) as the $\mathrm{X}_{\mathrm{C},i}$-axis, $\mathrm{Y}_{\mathrm{C},i}$-axis, and $\mathrm{Z}_{\mathrm{C},i}$-axis, whose origin and $\mathrm{Y}_{\mathrm{C},i}$-axis are the $i$th camera's location and orientation, respectively. Only horizontal deflection relative to the MCS is considered for each CCS, which means the $\mathrm{Z}_{\mathrm{C},i}$-axis is upward and is also perpendicular to the traffic lane plane, $i=1,2,\cdots,C$. We further define an RSU coordinate system (RCS) as the $\mathrm{X}_{\mathrm{R}}$-axis, $\mathrm{Y}_{\mathrm{R}}$-axis, and $\mathrm{Z}_{\mathrm{R}}$-axis, where the origin, $\mathrm{Y}_{\mathrm{R}}$-axis and $\mathrm{X}_{\mathrm{R}}$-axis are the RSU location, the right direction of traffic lane and the vertical direction of traffic lane respectively,  while $\mathrm{Z}_{\mathrm{R}}$-axis is upward and is perpendicular to the lane plane.

Note that, for an object, e.g., a vehicle or camera, we define the angle between the object orientation and the $\mathrm{Y}$-axis as the azimuth of the object in that specific coordinate system, as shown in Fig.~1. The azimuth is positive for clockwise rotation. For clarity, we utilize the subscript to distinguish the reference coordinate system of a coordinate value or an azimuth. For instance, $(x_{\mathrm{M}},y_{\mathrm{M}},z_{\mathrm{M}})$, $(x_{\mathrm{C},i},y_{\mathrm{C},i},z_{\mathrm{C},i})$ and $(x_{\mathrm{R}},y_{\mathrm{R}},z_{\mathrm{R}})$ denote the coordinate value under the MCS, the $i$th CCS, and the RCS, respectively. Moreover, $\theta_{\mathrm{M}}$, $\theta_{\mathrm{C},i}$ and $\theta_{\mathrm{R}}$ denote the azimuth under MCS, the $i$th CCS, and RCS, respectively.

The locations and the azimuths of the $i$th camera are fixed in MCS and are then denoted by $(x_{\mathrm{M}}^{i},y_{\mathrm{M}}^{i},z_\mathrm{M}^{i})$ and $\theta_\mathrm{M}^{i}$, $i=1,2,\cdots,C$ respectively. Without loss of generality, we assume the orientations $\theta_{\mathrm{M}}^{i}$ and the horizontal field of view (HFOV) of the equipped cameras can cover the horizontal view of 360 degrees around the MS, and the shooting scope of each camera does not overlap with each other. Assume there are $O_i$ vehicles in the $i$th image, $i=1,2,\cdots,C$, as shown in Fig.~2(a).\footnote{The value of $O_i$ could be different for different camera view} MS can apply 3D detection \cite{zliu} to detect the $j$th vehicles in the $i$th image, denoted as the $(i,j)$th vehicle, $j=1,2.,\cdots,O_i$. Moreover, 3D detection technique \cite{zliu} can also be used to obtain the length $l_{i,j}$, width $w_{i,j}$, height $h_{i,j}$, the center location $(x_{\mathrm{C},i}^{j},y_{\mathrm{C},i}^{j},z_{\mathrm{C},i}^{j})$ and the azimuth $\theta_{\mathrm{C},i}^{j}$ of the $(i,j)$th vehicle. All these detected vehicles' size/location/orientation parameters will be utilized latter to design the input feature of the DNN for the optimal beam alignment.

According to whether the MS knows its location, we separately discuss the following two cases:

\subsection{MS Location Is Known by Itself}
We propose the VBALA that contains following three key steps:
\begin{itemize}
    \item[1)]
     Transform the coordinates and azimuths of the vehicles from CCSs to the RCS.
    \item[2)]
    Perform gird quantization for the coordinates and azimuths of the vehicles under RCS to obtain a vehicle distribution feature (VDF).
    \item[3)]
    Design a VDF based beam alignment DNN (VDBAN) to infer the optimal beam pair from the VDF and MS location.
\end{itemize}

Specifically, we obtain the coordinates $(x^{i,j}_{\mathrm{M}},y^{i,j}_{\mathrm{M}},z^{i,j}_{\mathrm{M}})$ and azimuth $\theta_{\mathrm{M}}^{i,j}$ of the $(i,j)$th vehicle through the coordinate transformation between the MCS and the $i$th CCS:
\begin{equation}
\begin{aligned}
\left[
\begin{matrix}
  x^{i,j}_{\mathrm{M}} \\
  y^{i,j}_{\mathrm{M}} \\
  z^{i,j}_{\mathrm{M}}
\end{matrix}\right]&=\left[\begin{matrix}
\cos(\theta_{\mathrm{M}}^{i}) & \sin(\theta_{\mathrm{M}}^{i}) & 0 \\
-\sin(\theta_{\mathrm{M}}^{i}) & \cos(\theta_{\mathrm{M}}^{i}) & 0 \\
0 & 0 & 1
\end{matrix}\right]
\left[
\begin{matrix}
  x_{\mathrm{C},i}^{j} \\
  y_{\mathrm{C},i}^{j} \\
  z_{\mathrm{C},i}^{j}
\end{matrix}\right]+
\left[
\begin{matrix}
  x_{\mathrm{M}}^{i} \\
  y_{\mathrm{M}}^{i} \\
  z_{\mathrm{M}}^{i}
\end{matrix}\right],
\\
\theta_{\mathrm{M}}^{i,j}&=\theta_{\mathrm{M}}^{i}+\theta_{\mathrm{C},i}^{j}, j=1,2,\cdots,O_i, i=1,2,\cdots,C.
\end{aligned}
\end{equation}
Since the axes of MCS and RCS are parallel to each other, we can further get the coordinates $(x^{i,j}_{\mathrm{R}},y^{i,j}_{\mathrm{R}},z^{i,j}_{\mathrm{R}})$ and azimuth $\theta_{\mathrm{R}}^{i,j}$ of the $(i,j)$th vehicle from $(x_{\mathrm{R}}^{i,j},y_{\mathrm{R}}^{i,j},z_{\mathrm{R}}^{i,j})=(x_{\mathrm{M}}^{i,j}+x_{\mathrm{R}}^{\mathrm{MS}},y_{\mathrm{M}}^{i,j}+y_{\mathrm{R}}^{\mathrm{MS}},z_{\mathrm{M}}^{i,j}+z_{\mathrm{R}}^{\mathrm{MS}})$ and $\theta_{\mathrm{R}}^{i,j}=\theta_{\mathrm{M}}^{i,j}$, where $(x_{\mathrm{R}}^{\mathrm{MS}},y_{\mathrm{R}}^{\mathrm{MS}},z_{\mathrm{R}}^{\mathrm{MS}})$ are the MS's coordinate in RCS.

Let us then divide $\mathrm{X}_{\mathrm{R}}$-$\mathrm{Y}_{\mathrm{R}}$ plane of RCS into many grids with equal size, while denote the length and width of each grid as $L_{\mathrm{G}}$ and $W_{\mathrm{G}}$, respectively. We obtain those grids that intersect with the traffic lane, as shown in Fig. 2(b), and denote the number of the obtained grids as $G$. The area of the $g$th grid can be expressed as $\bm{\mathcal{G}}_g=\{(x_{\mathrm{R}},y_{\mathrm{R}},z_{\mathrm{R}})\ |\ i^{\mathrm{X}}_g L_{\mathrm{G}} \leq x_{\mathrm{R}}< (i^{\mathrm{X}}_g+1)L_{\mathrm{G}}, i^{\mathrm{Y}}_g W_{\mathrm{G}} \leq y_{\mathrm{R}}< (i^{\mathrm{Y}}_g+1)W_{\mathrm{G}},z_{\mathrm{R}}=0\}$, $g=1,2,\cdots,G$, where $(i^{\mathrm{X}}_g L_{\mathrm{G}}, i^{\mathrm{Y}}_g W_{\mathrm{G}}, 0)$ is a vertex of the grid and both $i^{\mathrm{X}}_g$ and $i^{\mathrm{Y}}_g$ are integers. Denote $\bm{\mathcal{V}}_g$ as the index set of the vehicles whose $\mathrm{X}_{\mathrm{R}}$-$\mathrm{Y}_{\mathrm{R}}$ plane locations are contained in $\bm{\mathcal{G}}_{g}$, with $\bm{\mathcal{V}}_g=\{(i,j)\ |\ (x^{i,j}_{\mathrm{R}},y^{i,j}_{\mathrm{R}},0)\in \bm{\mathcal{G}}_{g}, j=1,2,\cdots,O_i, i=1,2,\cdots,C\}$. Denote the maximum length, width and height of the vehicles in $\bm{\mathcal{V}}_g$ as $l_{\mathrm{max},g}$, $w_{\mathrm{max},g}$ and $h_{\mathrm{max},g}$, respectively. We then normalize $l_{\mathrm{max},g}$, $w_{\mathrm{max},g}$ and $h_{\mathrm{max},g}$ by the maximum vehicle length $L_{\mathrm{max}}$, width $W_{\mathrm{max}}$ and height $H_{\mathrm{max}}$ of all possible vehicle types respectively, and obtain $l^{\mathrm{N}}_{\mathrm{max},g}=\frac{l_{\mathrm{max},g}}{L_{\mathrm{max}}}$, $w^{\mathrm{N}}_{\mathrm{max},g}=\frac{w_{\mathrm{max},g}}{W_{\mathrm{max}}}$ and $h^{\mathrm{N}}_{\mathrm{max},g}=\frac{h_{\mathrm{max},g}}{H_{\mathrm{max}}}$. The average value $\theta_{\mathrm{R}}^{g}$ of the azimuths of the vehicles in $\bm{\mathcal{V}}_g$ can be computed as $\theta_{\mathrm{R}}^{g}=\frac{1}{\mathrm{Card}(\bm{\mathcal{V}}_g)}\sum_{(i,j)\in \bm{\mathcal{V}}_g}\theta_{\mathrm{R}}^{i,j}$.

Then, we design a VDF that can represent the vehicle distribution relative to RSU. The VDF is defined as a $G\times 4$ dimensional matrix $\bm{F}\in \mathbb{R}^{G\times 4}$, and the $g$th row of $\bm{F}$ is set as $[l^{\mathrm{N}}_{\mathrm{max},g}, w^{\mathrm{N}}_{\mathrm{max},g}, h^{\mathrm{N}}_{\mathrm{max},g}, \theta_{\mathrm{R}}^{g}]$. When one grid does not contain vehicles, the corresponding row of $\bm{F}$ will be set as a zero vector. Thus, the VDF $\bm{F}$ can reflect the distribution of MS's surrounding vehicles.\footnote{Although the grid quantization operation only uses one cubic block to represent all the vehicles contained in a grid, one will see in the later simulation that the proposed VDF can still help provide better beam alignment than the existing methods.}

Next, we use transformer model \cite{Vaswani} and design a VDBAN that can fuse the VDF $\bm{F}$ and the known MS's location $(x_{\mathrm{R}}^{\mathrm{MS}},y_{\mathrm{R}}^{\mathrm{MS}},z_{\mathrm{R}}^{\mathrm{MS}})$, as shown in the Fig.~2(c). The self-attention module of transformer can conveniently perform the information exchange between data with different modalities \cite{JLu}-\cite{WKim}. We here adopt the TransFuser model architecture \cite{Prakash}. Since $\mathrm{z}_{\mathrm{R}}^{\mathrm{MS}}$ is constant, the input features of VDBAN include the plane coordinates $[x_{\mathrm{R}}^{\mathrm{MS}},y_{\mathrm{R}}^{\mathrm{MS}}]$ and the VDF $\bm{F}$. The matrix $\bm{F}$ will be input into several 1-dimensional (1-D) convolution layers and fully connected (FC) layers to obtain the vector $\bm{f}\in \mathbb{R}^{1\times D}$, and $[x_{\mathrm{M}},y_{\mathrm{M}}]^{\mathrm{T}}$ will also be fed into FC layers to obtain the vector $\bm{u}\in \mathbb{R}^{1\times D}$.

Then, the self-attention module will utilize the trainable weight matrices $\bm{W}_{\mathrm{Q}}\in \mathbb{R}^{D\times d}$, $\bm{W}_{\mathrm{K}}\in \mathbb{R}^{D\times d}$ and $\bm{W}_{\mathrm{V}}\in \mathbb{R}^{D\times d}$ to generate the \emph{queries}, the \emph{keys} and the \emph{values} for $\bm{u}$ and $\bm{f}$ respectively, whereas the \emph{queries}, the \emph{keys} and the \emph{values} of $\bm{u}$ are then given by $\bm{q}_{\mathrm{u}}=\bm{u}\bm{W}_{\mathrm{Q}}$, $\bm{k}_{\mathrm{u}}=\bm{u}\bm{W}_{\mathrm{K}}$ and $\bm{v}_{\mathrm{u}}=\bm{u}\bm{W}_{\mathrm{V}}$ respectively. Similarly, $\bm{q}_{\mathrm{f}}=\bm{f}\bm{W}_{\mathrm{Q}}$, $\bm{k}_{\mathrm{f}}=\bm{f}\bm{W}_{\mathrm{K}}$ and $\bm{v}_{\mathrm{f}}=\bm{f}\bm{W}_{\mathrm{V}}$ are the \emph{queries}, the \emph{keys} and the \emph{values} of $\bm{f}$ respectively. The dot products between each \emph{query} and all \emph{keys} are input into the Softmax function to determine the weights for all the \emph{values}, while the weighted sum of the \emph{values} is used as the layer output corresponding to the \emph{query}. Specifically, the output of self-attention module of $\bm{u}$ and $\bm{f}$ are denoted by $\bm{u}^{'}$ and $\bm{f}^{'}$ respectively, where
\begin{equation}
\begin{aligned}
&\left[\begin{matrix}
  \bm{u}^{'} \\
  \bm{f}^{'}
\end{matrix}\right]=\mathrm{Softmax}(\frac{1}{\sqrt{d}}\bm{Q}\bm{K}^{\mathrm{T}})\bm{V},\\
&\bm{Q}=\left[\begin{matrix}
  \bm{q}_{\mathrm{u}} \\
  \bm{q}_{\mathrm{f}}
\end{matrix}\right],\ \bm{K}=\left[\begin{matrix}
  \bm{k}_{\mathrm{u}} \\
  \bm{k}_{\mathrm{f}}
\end{matrix}\right],\ \bm{V}=\left[\begin{matrix}
  \bm{v}_{\mathrm{u}} \\
  \bm{v}_{\mathrm{f}}
\end{matrix}\right].
\end{aligned}
\end{equation}

We next utilize the \emph{multi-head} attention mechanism \cite{Vaswani} to enhance the representation ability of Transformer. For \emph{multi-head} attention, denote $h$ as the number of heads. We can utilize $h$ groups of the attention weight matrices $(\bm{W}_{\mathrm{Q}},\bm{W}_{\mathrm{K}},\bm{W}_{\mathrm{V}})$ to obtain different self-attention module outputs $(\bm{u}^{'}_1, \bm{f}^{'}_1)$, $(\bm{u}^{'}_2, \bm{f}^{'}_2)$, $\cdots$, $(\bm{u}^{'}_h, \bm{f}^{'}_h)$ according to equation (6). All the outputs of multi-head are concatenated and are mapped to the $D$-dimensional vector $\bm{u}_{\mathrm{o}}$ and $\bm{v}_{\mathrm{o}}$ by a linear mapping $\bm{W}_{\mathrm{O}}\in \mathbb{R}^{d*h\times D}$, i.e.,
\begin{equation}
\left[\begin{matrix}
  \bm{u}_{\mathrm{o}} \\
  \bm{f}_{\mathrm{o}}
\end{matrix}\right]=\left[\begin{matrix}
                           \mathrm{Concat}(\bm{u}^{'}_1,\bm{u}^{'}_2,\cdots,\bm{u}^{'}_h) \\
                           \mathrm{Concat}(\bm{f}^{'}_1,\bm{f}^{'}_2,\cdots,\bm{f}^{'}_h)
                         \end{matrix}\right]\bm{W}_{\mathrm{O}}.
\end{equation}

The self-attention module and the feed-forward network formulate a transformer encoder and both integrated with the residual connection \cite{HeKai}. The outputs of the transformer encoder are two $D$-dimensional vectors that are obtained by feeding $\bm{u}+\bm{u}_{\mathrm{o}}$ and $\bm{f}+\bm{f}_{\mathrm{o}}$ into the same feed-forward network with two FC layers respectively. The obtained outputs of the transformer encoder will then be repeatedly fed into the FC layers and the transformer encoder module. The two outputs from the final transformer encoder module will be added up and then be input into a multi-layer perception (MLP) network to generate the output of VDBAN. Denote the set of all possible beam pairs from codebooks $\bm{\mathcal{W}}_{\mathrm{B}}$ and $\bm{\mathcal{W}}_{\mathrm{U}}$ as $\bm{\mathcal{W}}_{\mathrm{P}}=\{(\bm{w}_{\mathrm{B},\mathrm{1}},\bm{w}_{\mathrm{U},\mathrm{1}}),(\bm{w}_{\mathrm{B},1},\bm{w}_{\mathrm{U},2}),\cdots,(\bm{w}_{\mathrm{B},N^{\mathrm{CB}}_{\mathrm{B}}},\bm{w}_{\mathrm{U},N^{\mathrm{CB}}_{\mathrm{U}}})\}$, where $\mathrm{Card}(\bm{\mathcal{W}}_{\mathrm{P}})=N^{\mathrm{CB}}_{\mathrm{B}}N^{\mathrm{CB}}_{\mathrm{U}}$. The output of VDBAN is the index of the optimal beam pair in set $\bm{\mathcal{W}}_{\mathrm{P}}$. With the predicted optimal beam pair index from VDBAN, the optimal transmit and receive beam can be indexed in $\bm{\mathcal{W}}_{\mathrm{B}}$ and $\bm{\mathcal{W}}_{\mathrm{U}}$ respectively. Note that, the proposed beam alignment method is performed at MS and the index of the optimal transmit beam is fed back to RSU.

\begin{figure}[t]
\centering
\includegraphics[width=1\textwidth]{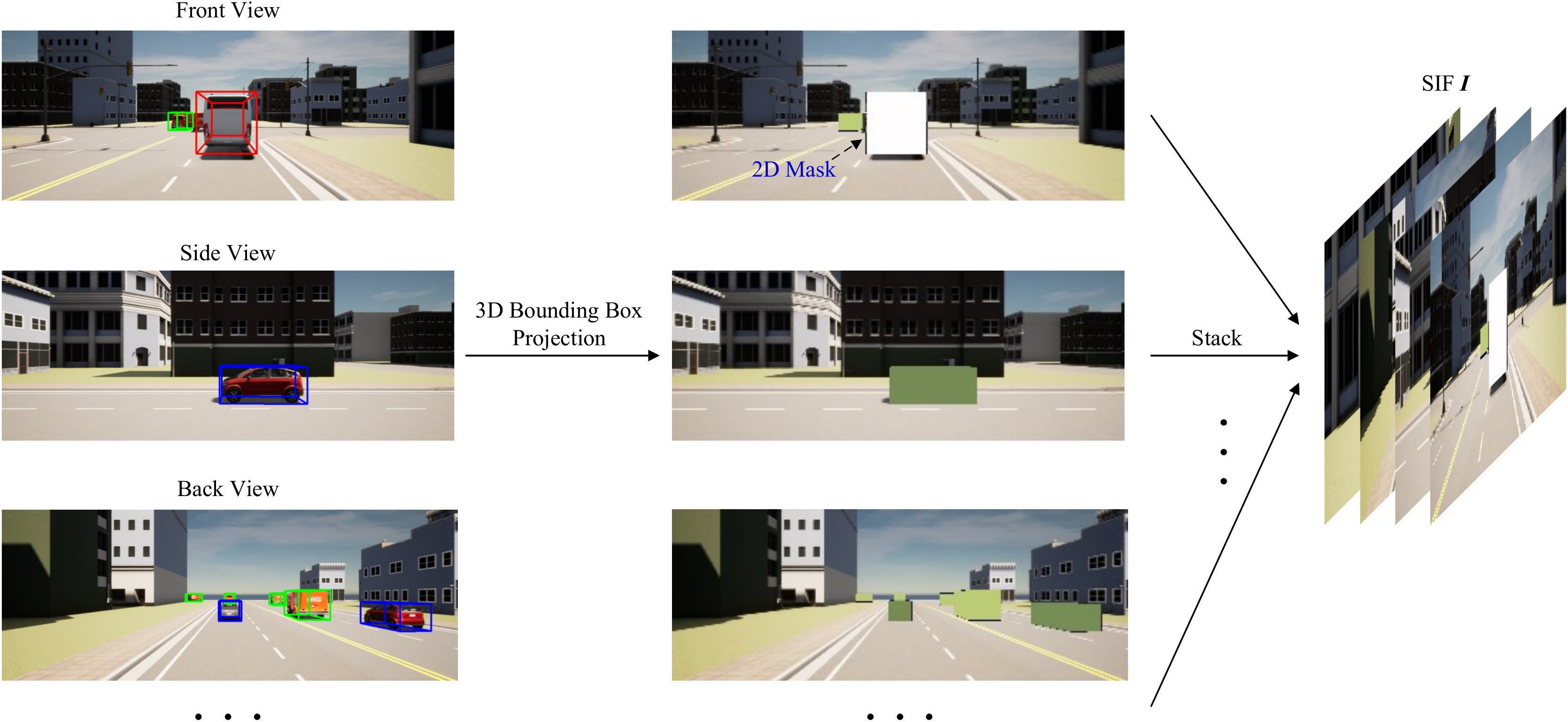}
\caption{The design of SIF for VBALU.}
\end{figure}
\subsection{MS Location Is Not Known}
When the MS's location information cannot be obtained accurately and immediately, the VDBAN cannot be applied or would present severe performance loss. In this case, we consider to exploit the background information in the images, since the appearance and the distribution information of surrounding buildings can effectively reflect the MS location information \cite{Xu2}. Furthermore, the distribution of the pixels of vehicles in the images, i.e., the foreground information of images, can also intuitively reflect the vehicle distribution relative to the MS. For example, the orange van shown in the left part of the image taken at front view in Fig.~2(a) can approximately indicate the existence of a van on the left-front of the MS. Thus, the images can indicate both the MS's location information and the scattering characteristics of surrounding vehicles, and thereby can be used to infer the best beam pair without the MS's location.

However, the size of pixel area of the vehicle cannot represent the vehicle’s actual size and distance from MS. For instance, the vehicle with small size can still look very large in the image when the vehicle is close to the camera. This is the inherent defect of 2D visual perception of the ordinary camera, but can be well handled by the 3D perception of Lidar or RGB-D camera. Moreover, compared with the size/location/orientation of vehicles, the color information of pixels of the vehicles is unimportant and is redundant for beam alignment, as the vehicle color may have almost no impact on the channel propagation but will increase the learning difficulty of DNN.

Hence, we design a scene image feature (SIF) by converting the vehicle color information to the vehicle size information. Specifically, we replace the three color components of the pixels of each vehicle with the length, width and height of the vehicle. For the $(i,j)$th vehicle, we utilize $(-255\frac{l_{i,j}}{L_{\mathrm{max}}},-255\frac{w_{i,j}}{W_{\mathrm{max}}},-255\frac{h_{i,j}}{H_{\mathrm{max}}})$ as the RGB channel values for all pixels of the vehicle, where the minus sign is used to distinguish the pixels of vehicles from the pixels of the scene background. We estimate the 3D bounding box of the vehicle by 3D detection. As Fig.~3 shows, we project the 3D bounding box onto the image to generate a compact 2D mask. The pixels of the vehicle are considered to be the pixels contained in the 2D mask. In fact, the vehicles pixels can be extracted more precisely by the foreground or background segmentation method \cite{Akilan}. However, the 2D mask projected by 3D bounding box is utilized here for simplicity. Although the adopted 2D mask is slightly coarse to cover the vehicle pixels, the effectiveness of the VBALU can still be demonstrated through the simulation.

When the vehicles in all the $C$ images are masked with the size information, we concatenate the $C$ images into a $3C$-channel image to formulate the SIF $\bm{I}$. Specifically, the SIF $\bm{I}\in \mathbb{R}^{f_{\mathrm{H}}\times f_{\mathrm{W}}\times 3C}$ is defined as a 3D matrix, where $f_{\mathrm{H}}$ and $f_{\mathrm{W}}$ are the height and the width of the image respectively. The $i$th image with 2D masks is used as $\bm{I}_{[:,:,(3c-2):3c]}$, $i=1,2,\cdots,C$. Then, we design an SIF based beam alignment DNN (SIBAN) that adopts the widely-used ResNet architecture, such as the well-known ResNet-34 or ResNet-50 architecture \cite{HeKai}, to predict the optimal beam pair from SIF. The input of the SIBAN is $\bm{I}$ and the output is the same as VDBAN.

\emph{\textbf{Remark}}: It is worth noting that though VBALA requires the MS's location that is unnecessary for VBALU, VBALA has better environmental adaptability and generalization than VBALU. Specifically, the beam alignment performance of VBALA is almost unaffected by the surrounding environmental buildings, as the main scatters are the vehicles in traffic lane. However, since the VBALU need to infer MS's location information from the background information of images, the beam alignment performance of VBALU will be related to the spatial/color characteristics of environmental buildings.

\section{Vision Based BCT Prediction Method}
\begin{figure}[t]
\centering
\includegraphics[width=0.75\textwidth]{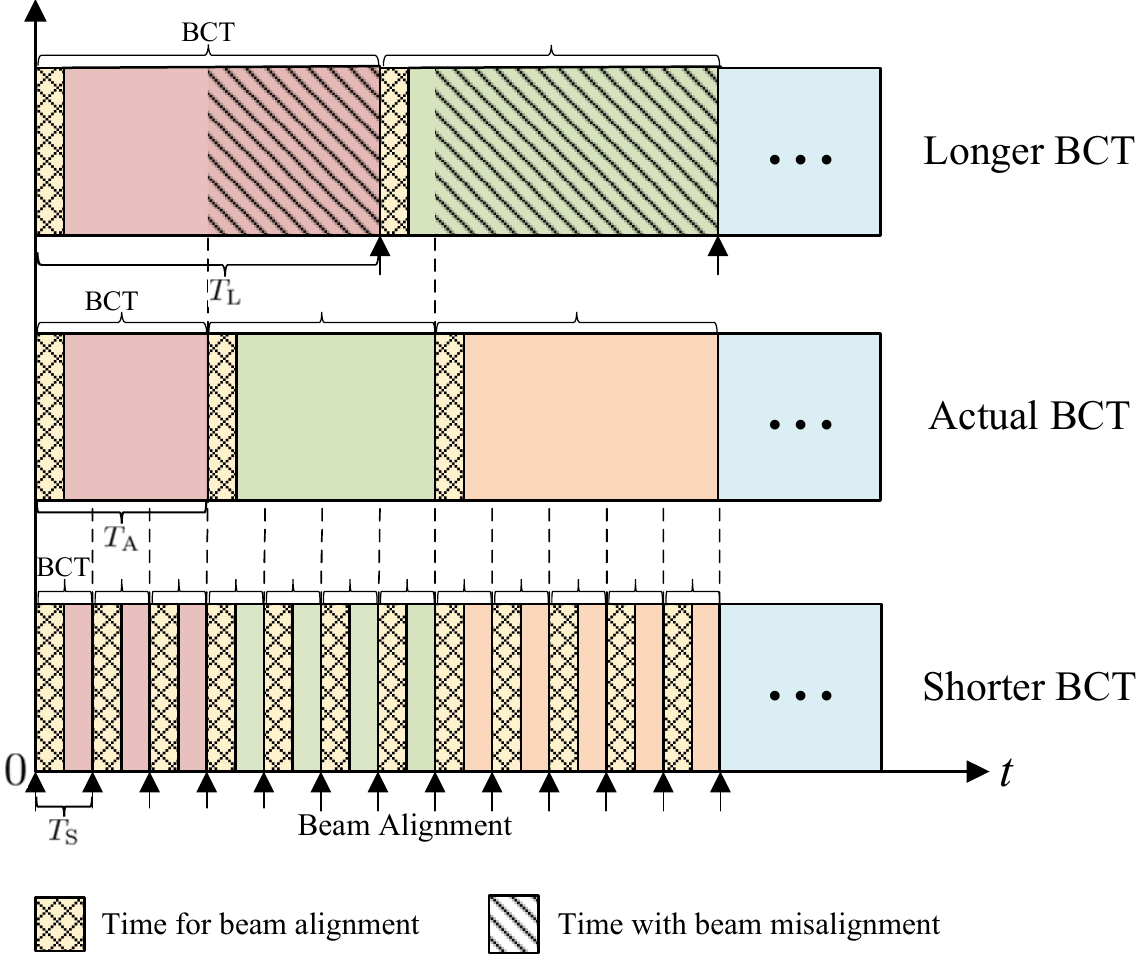}
\caption{Trade-off between the length of BCT and the transmission rate.}
\end{figure}
\begin{figure*}[t]
\centering
\includegraphics[width=1\textwidth]{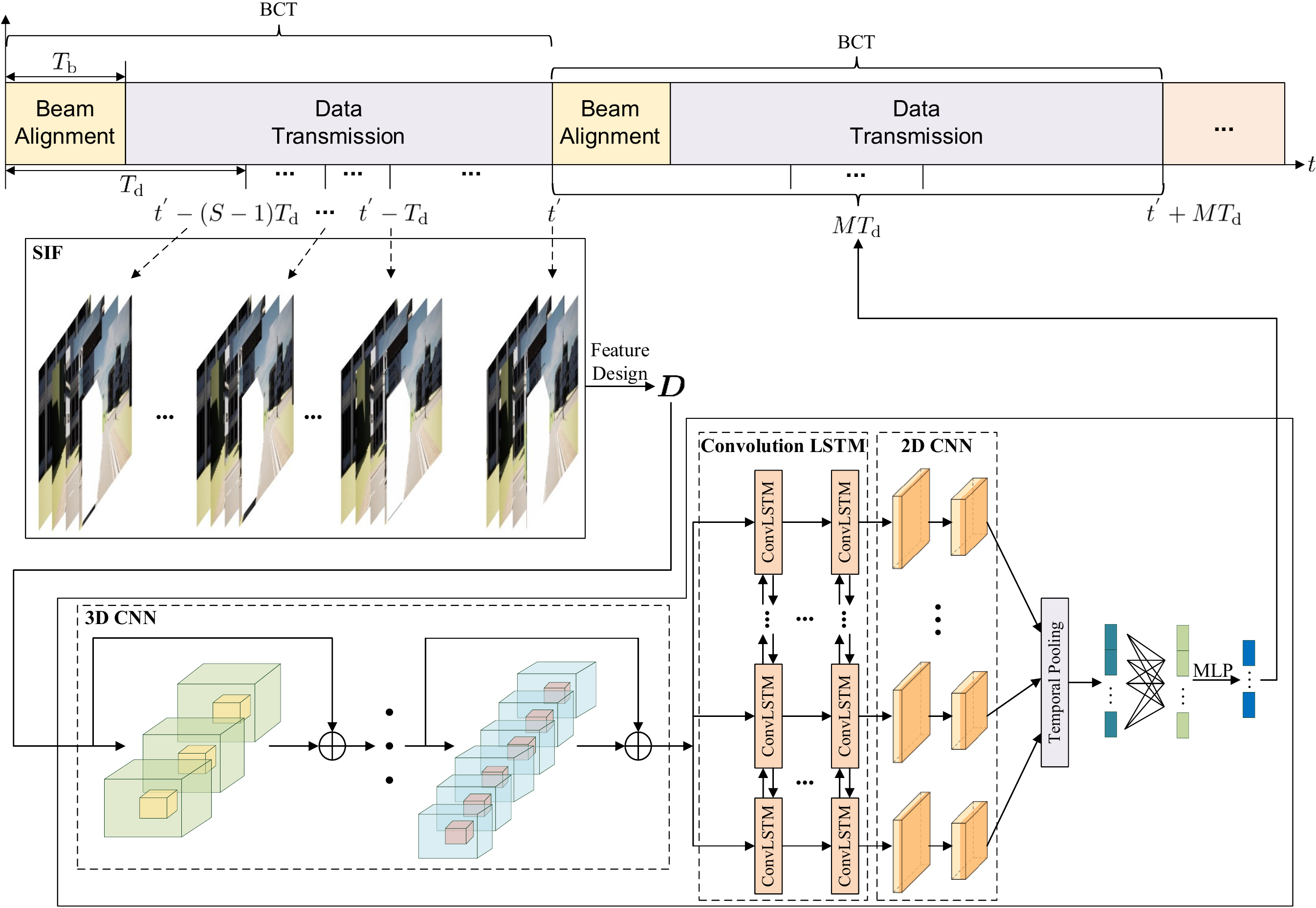}
\caption{The diagram of the proposed VPBCT.}
\end{figure*}

The accurate acquisition of BCT, i.e., the duration of the optimal beam pair, is crucial to improve the transmission rate. As Fig.~4 shows, the estimated BCT $T_{\mathrm{L}}$ that is longer than the actual BCT $T_{\mathrm{A}}$ can cause severe beam misalignment during some time periods, such as the time period $[T_{\mathrm{A}},T_{\mathrm{L}}]$ and thereby results in lower transmission rate than that can be achieved by the actual BCT. When the estimated BCT $T_{\mathrm{S}}$ is shorter than the actual BCT $T_{\mathrm{A}}$, the beam alignment will be frequently performed, as Fig.~4 shows the beam alignment is repeated for 3 times during the actual BCT period $[0,T_{\mathrm{A}}]$. Thus, the transmission rate will also decrease due to the large time overhead for beam alignment.

Specifically, the achievable transmission rate is affected by two factors: the beam alignment accuracy and the beam alignment time. Denote the total time for communications as $T_{\mathrm{total}}$, and denote the total time for beam alignment as $T_{\mathrm{align}}$. Moreover, the average transmission rate achieved by the beam alignment for the time $T_{\mathrm{total}}-T_{\mathrm{align}}$ without beam alignment overhead is denoted as $R_{\mathrm{ave}}$, and $R_{\mathrm{ave}}$ is positively correlated with the beam alignment accuracy. Thus, the actual average transmission rate considering the time overhead of beam alignment can be expressed as $(1-\frac{T_{\mathrm{align}}}{T_{\mathrm{total}}})R_{\mathrm{ave}}$. When the estimated BCT is longer than the actual BCT, $T_{\mathrm{align}}$ can become shorter, but $R_{\mathrm{ave}}$ may severely decrease due to the beam misalignment. When the estimated BCT is shorter than the actual BCT, the $R_{\mathrm{ave}}$ may be optimal, but the $T_{\mathrm{align}}$ will increase to degrade the actual average transmission rate. Thus, the accurate BCT is expected to obtain to realize the optimal trade-off between $R_{\mathrm{ave}}$ and $T_{\mathrm{align}}$.

The BCT is mainly affected by the location, size, moving direction and speeds of the MS and the surrounding scatters. Conventionally, the BCT can only be estimated under some LOS scenarios ordinarily, such as the high speed train communication and unmanned aerial vehicle communication, from the MS velocity and the size of beam coverage area \cite{LYang}-\cite{VVa}. For the complex environment with dynamic scatters and blockages, the traditional beam alignment strategy generally adopts the fixed BCT that is not guaranteed to be accurate \cite{Ahmed1}-\cite{SJiang},\cite{DZhang}. As the consecutively perceived scene images at MS can effectively present the exact spatial information and moving characteristics of the environmental objects, we use DNN to predict the accurate BCT from the scene images.

The diagram of the proposed VPBCT method is shown in Fig.~5. Denote $T_{\mathrm{d}}$ as the shooting interval of the camera, i.e., the reciprocal of the camera's frame rate, and denote $T_{\mathrm{b}}$ as the time for beam alignment during one BCT. During each BCT, the communication system will first perform beam alignment to estimate the optimal beam pair and then utilize the estimated beam pair for data transmission. Since the actual BCT is a continuous variable, we discretize BCT with duration $T_{\mathrm{d}}$ to reduce the learning difficulty of the DNN. Without loss of generality, we express the BCT of the optimal beam pair at moment $t^{'}$ as $MT_{\mathrm{d}}$, where $M$ is an integer with $M\geq 1$. Thus, the minimum BCT that may be adopted by system is $T_{\mathrm{d}}$, and the time $T_{\mathrm{b}}$ for beam alignment is assumed to be less than the minimum BCT, i.e., $T_{\mathrm{b}}<T_{\mathrm{d}}$, for simplicity.

We design a scene image based BCT prediction DNN (SIBPN) that adopts the 3D CNN integrated with the ConvLSTM (3D CNN-LSTM) model \cite{LZhang} to predict $M$. As shown in Fig.~5, the SIBPN is formed by connecting 3D CNN, ConvLSTM and 2D CNN module in series. The 3D CNN module includes several 3D CNN layers with residual connection, the ConvLSTM module consists of the bidirectional ConvLSTM layers, and the 2D CNN module consists of the ordinary 2D CNNs. The ConvLSTM is designed by replacing the fully connected operators of the traditional LSTM with the convolution operator to obtain the spatial information in the images \cite{xing}.

We utilize the images that are taken by the $C$ cameras at the MS to generate the input feature $\bm{D}$ of the SIBPN. Specifically, $\bm{D}\in \mathbb{R}^{S\times f_{\mathrm{H}}\times f_{\mathrm{W}}\times 3C}$ is a 4-dimensional matrix, and its $p$th row is the SIF generated by the $C$ images taken at moment $t^{'}-(S-p)T_{\mathrm{d}}$, $p=1,2,\cdots,S$. The input feature $\bm{D}$ is fed into the 3D CNN module to obtain a 4D tensor $\bm{D}^{'}\in \mathbb{R}^{S^{'}\times f_{\mathrm{H}}^{'}\times f_{\mathrm{W}^{'}}\times 3C^{'}}$. Then, $\bm{D}^{'}$ is split into a sequential image data with $S^{'}$ time steps, i.e., $\bm{D}^{'}_{[1,:,:,:]},\bm{D}^{'}_{[2,:,:,:]},\cdots,\bm{D}^{'}_{[S^{'},:,:,:]}$. The sequential image data is fed into the ConvLSTM module to extract both the temporal and spatial characteristics of the environmental objects. The output of ConvLSTM of each time step will be fed into a 2D CNN respectively for dimension reduction. Finally, the average of the output vectors of all 2D CNNs are calculated for temporal pooling, and this average vector is fed into several FC layers to produce the output of the SIBPN. Since $M$ is assumed as an integer, the BCT prediction problem can be regarded as a classification problem. We here assume a maximum possible length of BCT as $M_{\mathrm{max}}T_{\mathrm{d}}$. Thus, $M$ must be in the set $\bm{\mathcal{B}}=\{1,2,\cdots,M_{\mathrm{max}}\}$, and the output of the SIBPN is the index of the accurate BCT in set $\bm{\mathcal{B}}$.

Once $M$ is predicted by SIBPN, the communication system will implement beam alignment during the time period $[t^{'}, t^{'}+T_{\mathrm{b}}]$, and then keeps on using the optimal beam pair obtained from beam alignment during the time period $[t^{'}+T_{\mathrm{b}}, t^{'}+MT_{\mathrm{d}}]$. The next BCT prediction as well as the beam alignment will be performed at moment $t^{'}+MT_{\mathrm{d}}$.

\section{Simulation Results}

\begin{figure}
  \centering
\subfigure[]{
\begin{minipage}[t]{0.6\linewidth}
\centering
\includegraphics[height=60mm,width=90mm]{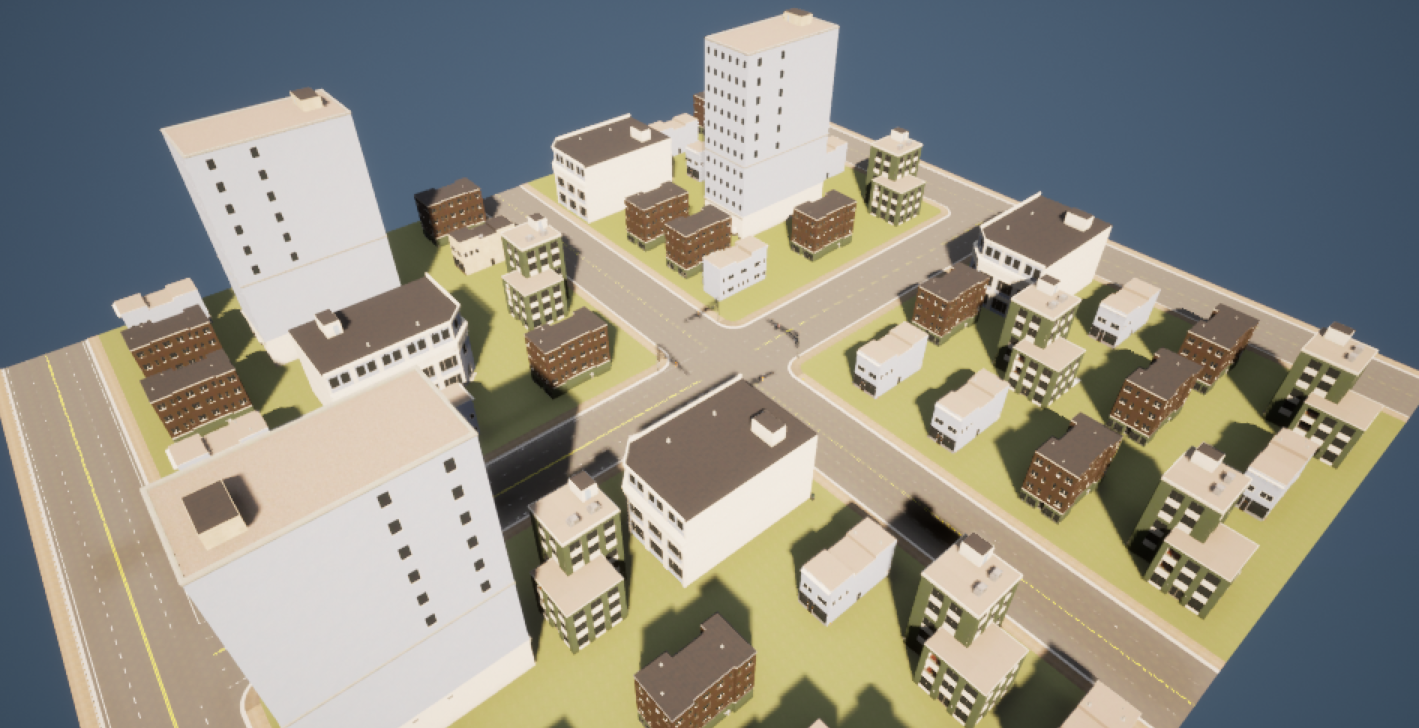}
%\caption{fig1}
\end{minipage}%
}%
\subfigure[]{
\begin{minipage}[t]{0.4\linewidth}
\centering
\includegraphics[width=55mm]{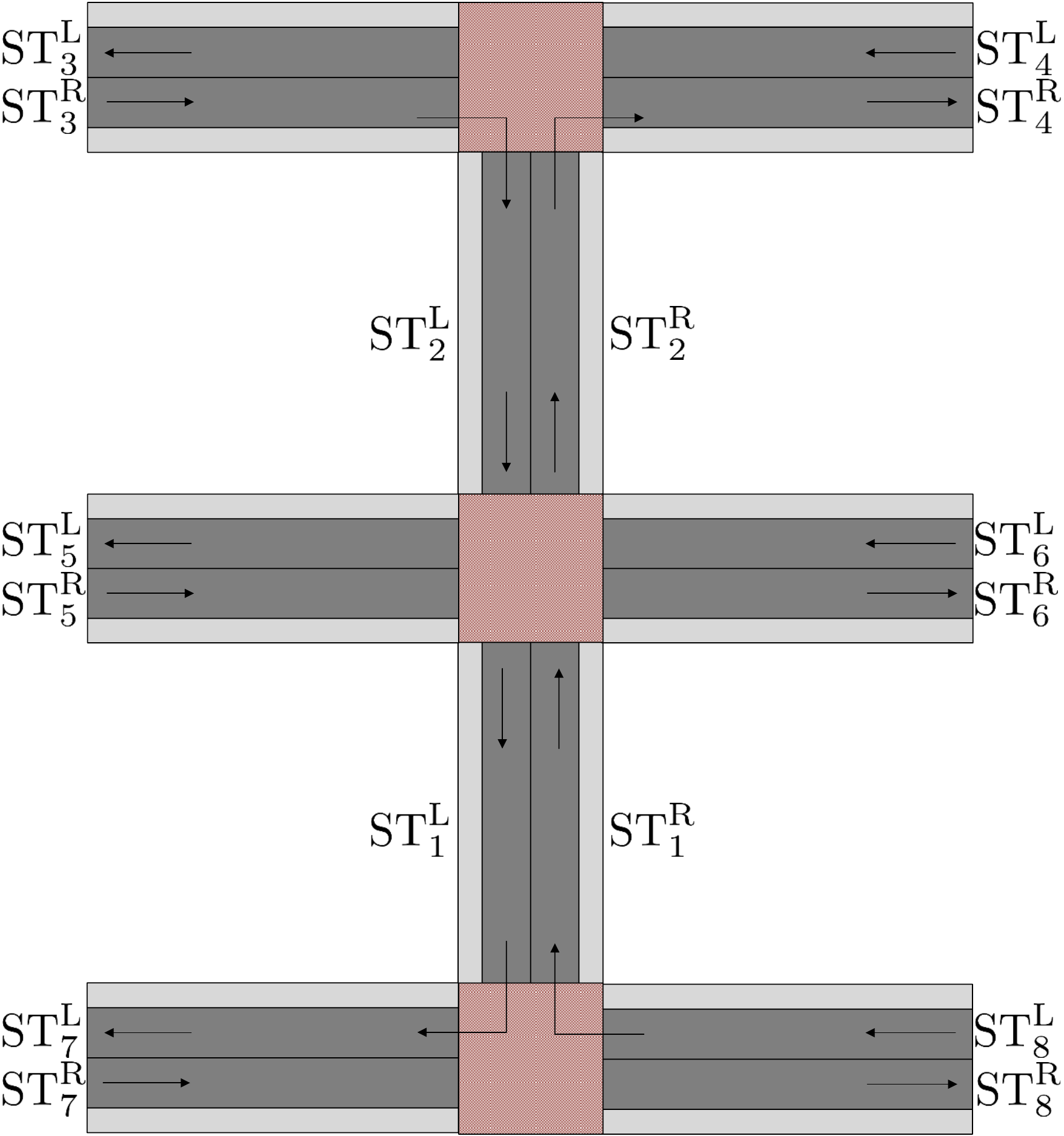}
%\caption{fig2}
\end{minipage}
}
\caption{(a) The constructed communication environment in CARLA. (b) The eight streets of the communication environment.}
\end{figure}

\begin{figure}[t]
\centering
\includegraphics[width=0.75\textwidth]{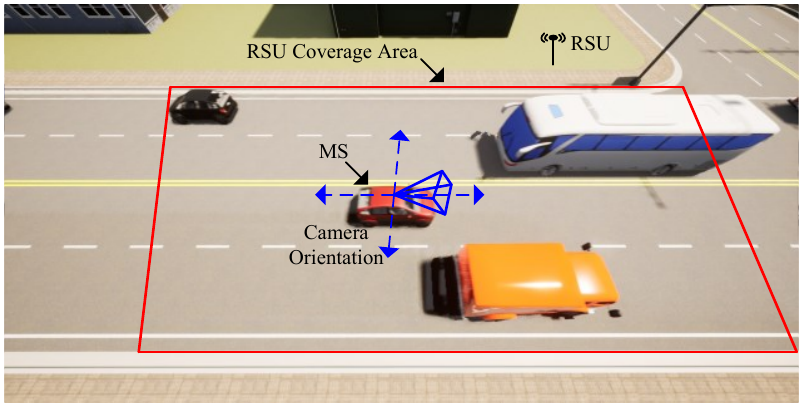}
\caption{The simulated RSU coverage area.}
\end{figure}
In this section, we verify the performance of VBALAA, VBALAU and VPBCT in a simulated traffic V2X environment. The images and channels are collected to construct the datesets of VDBAN, SIBAN and SIBPN.
\subsection{Simulation Setup}

\subsubsection{Environment Generation}
\begin{table}[t]
\centering
\caption{Vehicle Sizes For Simulation}
\begin{tabular}{|c|c|c|c|}
\hline
Type& Length/m& Width/m& Height/m\\
\hline
Car& 3.71& 1.79& 1.55\\
\hline
Van & 5.20& 2.61& 2.47\\
\hline
Bus & 11.08& 3.25& 3.33\\
\hline
\end{tabular}
\end{table}
We utilize the CARLA \cite{carla}, an autonomous driving simulation platform, to simulate the environment as well as vehicles. The CARLA could support to equip the vehicle with many kinds of sensors, such as camera, LIDAR and inertial measurement unit (IMU), etc., to obtain abundant sensory measurement data. Fig.~6(a) shows the constructed 3D communication environment model in CARLA. The environment model has eight streets. Denote the left and the right lane of the $r$th street as $\mathrm{ST}_{r}^{\mathrm{L}}$ and $\mathrm{ST}_{r}^{\mathrm{R}}$ respectively, $r=1,2,\cdots,8$, as shown in Fig.~6(b). The specified driving direction for each lane is also shown in Fig.~6(b).

The RSU is placed at the side of $\mathrm{ST}_1^{\mathrm{L}}$. The concerned RSU's coverage area is part of $\mathrm{ST}_1^{\mathrm{L}}$ and $\mathrm{ST}_1^{\mathrm{R}}$, which is 30 meters long and 15 meters wide, as shown in the Fig.~7. Three vehicle types including \emph{car type}, \emph{van type} and \emph{bus type} are adopted, whose sizes are listed in TABLE~III. To simulate traffic scenario, we first randomly generate 15 vehicles in $\mathrm{ST}_1^{\mathrm{L}}$, $\mathrm{ST}_2^{\mathrm{L}}$, $\mathrm{ST}_3^{\mathrm{R}}$ and also randomly generate 15 vehicles in $\mathrm{ST}_1^{\mathrm{R}}$, $\mathrm{ST}_2^{\mathrm{R}}$, $\mathrm{ST}_8^{\mathrm{L}}$ for the vehicle initialization. Here, we will give priority to generate a vehicle with the \emph{car type} in $\mathrm{ST}_1^{\mathrm{R}}$ and set the vehicle as MS. The types of all generated vehicles and the colors of the vehicles belonging to \emph{car type} are randomly determined. Then, we utilize the Simulation of Urban MObility (SUMO) \cite{sumo}, a traffic simulation software, to control the speed and moving trajectory of all the vehicles by the co-simulation interface of the CARLA.

\subsubsection{Scene Image Generation}
The locations of all $C$ cameras are set at $0.5\mathrm{m}$ above the roof center of MS, and the $C$ cameras have different orientations. When the MS is running through the RSU's coverage area, it keeps on taking images with the interval $T_{\mathrm{d}}$ until it leaves the area. The $C$ images taken at each moment form an image set, and the MS can collect a sequence of image sets within the RSU's coverage area.

\subsubsection{Channel Generation}
\begin{table}[t]
\centering
\caption{Critical Parameters of Wireless Insite for Ray Tracing}
\begin{tabular}{|c|c|}
\hline
Parameter& Value\\
\hline
Carrier Frequency& 28 $\mathrm{GHz}$\\
\hline
Propagation Model& X3D\\
\hline
Building Material& Concrete\\
\hline
Vehicle Material& Metal\\
\hline
Maximum Number of Reflections& 6\\
\hline
Maximum Number of Diffractions& 1\\
\hline
Maximum Paths Per Receiver Point& 5\\
\hline
\end{tabular}
\end{table}
We adopt the Wireless Insite \cite{remcom}, a ray tracing software, to simulate the channel. Note that the channel produced by ray tracing technique can have consistent spatial correlation with the communication environment. The setup parameters of Wireless Insite are listed as TABLE~IV. The ULA equipped at BS and MS are both set to be parallel with $\mathrm{ST}_1^{\mathrm{L}}$. The ULA of MS is equipped at $0.05\mathrm{m}$ above the roof center of MS. The ULA of RSU is equipped at $3\mathrm{m}$ above the ground. For each moment that MS take images, we will synchronize the environment and all vehicles of the CARLA into Wireless Insite. Specifically, as Fig.~8 shows, we ensure the environment model and the sizes/locations/orientations of vehicles in Wireless Insite are exactly the same as that in CARLA to produce the corresponding channel. Thus, the produced channel can match the image set taken at the same moment.

To reduce the computation overhead of channel generation, we convert the environment model and the vehicle models of CARLA into the simple cube-like models by ignoring the surface details of the models of CARLA, and utilize the cube-like models in Wireless Insite. Since the small surface details of CARLA models can only have a slight impact on the channel, this simplification for CARLA models will not affect the reliability of simulation.
\begin{figure}[t]
\centering
\includegraphics[width=0.75\textwidth]{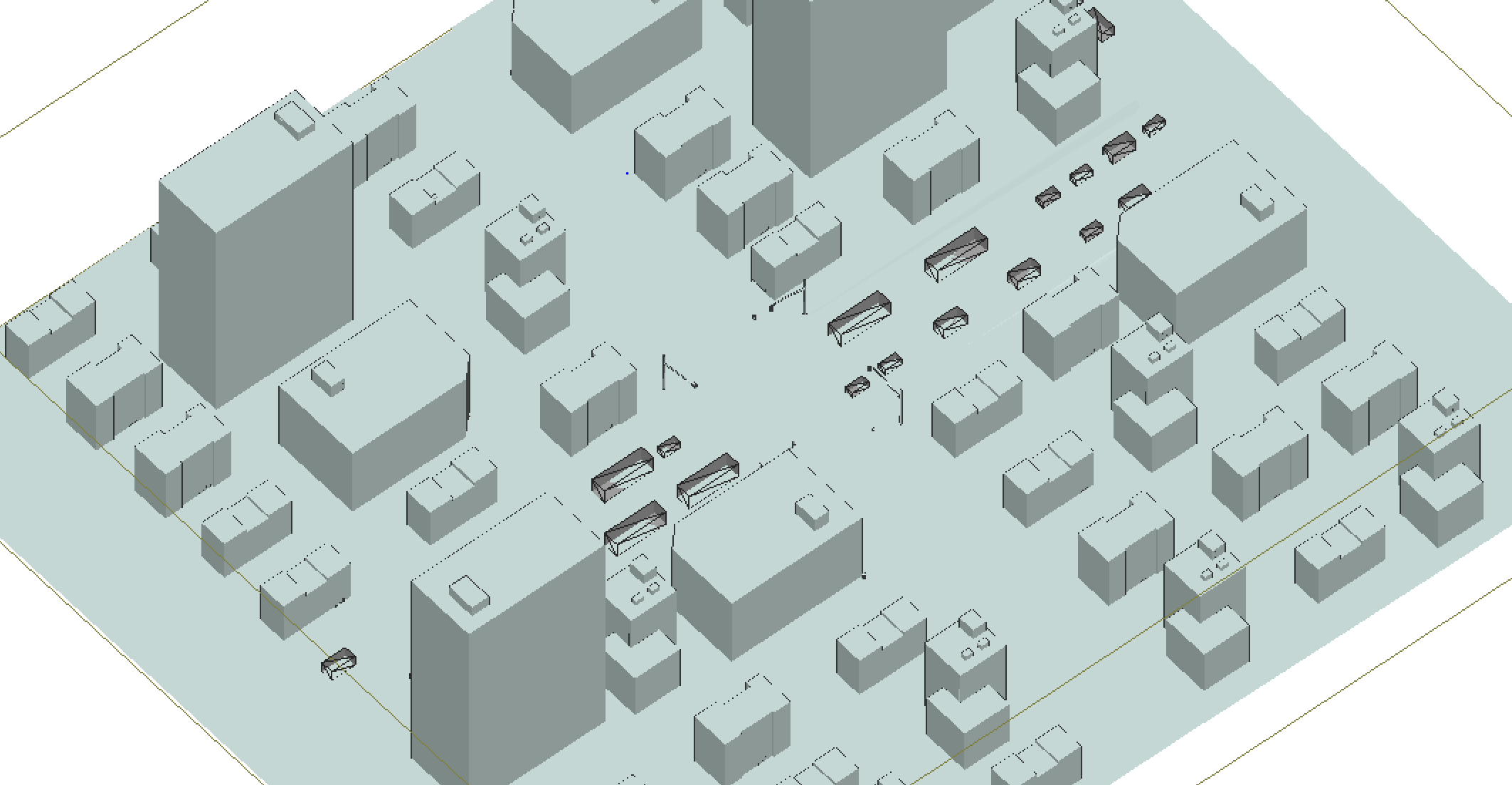}
\caption{The synchronization simulation in Wireless Insite.}
\end{figure}

\subsubsection{DNN Dataset Generation}
According to Section V.A-1) and Section V.A-2), the types and initial locations of generated vehicles are different with different vehicle initializations, and thus the image set sequences collected by MS will be different from each other. To obtain the datasets of VDBAN/SIBAN/SIBPN, we implement $Q$ different vehicle initializations and thereby acquire $Q$ different image set sequences. Denote the number of image sets contained in the $q$th image set sequence as $S_{q}$, and denote the $r$th image set of the $q$th image set sequence as $\bm{\mathcal{I}}_{q,r}$, $r=1,2,\cdots,S_q$, $q=1,2,\cdots,Q$. The channel corresponding to $\bm{\mathcal{I}}_{q,r}$ will be generated according to Section V.A-3), and the optimal beam pair index $b_{q,r}^{\mathrm{opt}}$ can be calculated from the channel. We also record the MS's location each shooting moment and denote the MS location as $\bm{p}_{q,r}$. Moreover, the VDF $\bm{F}_{q,r}$ can be generated from $\bm{\mathcal{I}}_{q,r}$ by the 3D detection method.

Thus, we can pair all the VDFs and MS locations with the corresponding optimal beam pair indexes and construct the sample set $\bm{\mathcal{S}}_{\mathrm{VDBAN}}=\{(\bm{F}_{q,r},\bm{p}_{q,r},b_{q,r}^{\mathrm{opt}})\ |\ r=1,2,\cdots,S_q, q=1,2,\cdots,Q\}$. Each element in $\bm{\mathcal{S}}_{\mathrm{VDBAN}}$ can be a training/validation/test sample of VDBAN. We divide the $\bm{\mathcal{S}}_{\mathrm{VDBAN}}$ into the training set $\bm{\mathcal{S}}_{\mathrm{VDBAN}}^{\bm{\mathcal{Q}}_{\mathrm{T}}}$, the validation set $\bm{\mathcal{S}}_{\mathrm{VDBAN}}^{\bm{\mathcal{Q}}_{\mathrm{V}}}$ and test set $\bm{\mathcal{S}}_{\mathrm{VDBAN}}^{\bm{\mathcal{Q}}_{\mathrm{E}}}$ of VDBAN, where $\bm{\mathcal{S}}_{\mathrm{VDBAN}}^{\bm{\mathcal{Q}}}=\{(\bm{F}_{q,r},\bm{p}_{q,r},b_{q,r}^{\mathrm{opt}})\ |\ r=1,2,\cdots,S_q, q\in \bm{\mathcal{Q}}\}$ and $\bm{\mathcal{Q}}=\bm{\mathcal{Q}}_{\mathrm{T}},\bm{\mathcal{Q}}_{\mathrm{V}},\bm{\mathcal{Q}}_{\mathrm{E}}$. Moreover, the $\bm{\mathcal{Q}}_{\mathrm{T}}$, $\bm{\mathcal{Q}}_{\mathrm{V}}$ and $\bm{\mathcal{Q}}_{\mathrm{E}}$ are the sets of the indices of image set sequences used for constructing training, validation and test set respectively, with $\mathrm{Card}(\bm{\mathcal{Q}}_{\mathrm{T}})+\mathrm{Card}(\bm{\mathcal{Q}}_{\mathrm{V}})+\mathrm{Card}(\bm{\mathcal{Q}}_{\mathrm{E}})=Q$. Denote the SIF generated from $\bm{\mathcal{I}}_{q,r}$ as $\bm{I}_{q,r}$. Similarly for the SIBAN, we can obtain the training set $\bm{\mathcal{S}}_{\mathrm{SIBAN}}^{\bm{\mathcal{Q}}_{\mathrm{T}}}$, the validation set $\bm{\mathcal{S}}_{\mathrm{SIBAN}}^{\bm{\mathcal{Q}}_{\mathrm{V}}}$ and the test set $\bm{\mathcal{S}}_{\mathrm{SIBAN}}^{\bm{\mathcal{Q}}_{\mathrm{E}}}$, where $\bm{\mathcal{S}}_{\mathrm{SIBAN}}^{\bm{\mathcal{Q}}}=\{(\bm{I}_{q,r},b_{q,r}^{\mathrm{opt}})\ |\ r=1,2,\cdots,S_q, q\in \bm{\mathcal{Q}}\}$ and $\bm{\mathcal{Q}}=\bm{\mathcal{Q}}_{\mathrm{T}},\bm{\mathcal{Q}}_{\mathrm{V}},\bm{\mathcal{Q}}_{\mathrm{E}}$.

To train the SIBPN, we should first determine the BCT of the optimal beam pair $b_{q,r}^{\mathrm{opt}}$. For simplicity, we assume the channel will not change during a shooting interval $T_{\mathrm{d}}$ to ensure the BCT will be the integral multiple of $T_{\mathrm{d}}$. We compare whether the element in $\{b_{q,r+r^{'}}^{\mathrm{opt}}\ |\ r^{'}=1,2,\cdots,S_{q}-r\}$ is the same as $b_{q,r}^{\mathrm{opt}}$ one by one to find the maximum duration of $b_{q,r}^{\mathrm{opt}}$. Specifically, if $b_{q,r}^{\mathrm{opt}}=b_{q,r+1}^{\mathrm{opt}}=\cdots=b_{q,r+M_{q,r}-1}^{\mathrm{opt}}\neq b_{q,r+M_{q,r}}^{\mathrm{opt}}$, then we can determine the BCT of $b_{q,r}^{\mathrm{opt}}$ as $M_{q,r}T_{\mathrm{d}}$. Next, $\bm{\mathcal{I}}_{q,r-S+1}, \bm{\mathcal{I}}_{q,r-S+2}, \cdots, \bm{\mathcal{I}}_{q,r}$ can be used to generate the SIBPN's input feature $\bm{D}_{q,r}$ for predicting $M_{q,r}$, $r=S,S+1,\cdots,S_q$. For SIBPN, we pair $\bm{D}_{q,r}$ and $M_{q,r}$ to construct the training set $\bm{\mathcal{S}}_{\mathrm{SIBPN}}^{\bm{\mathcal{Q}}_{\mathrm{T}}}$, the validation set $\bm{\mathcal{S}}_{\mathrm{SIBPN}}^{\bm{\mathcal{Q}}_{\mathrm{V}}}$, and the test set $\bm{\mathcal{S}}_{\mathrm{SIBPN}}^{\bm{\mathcal{Q}}_{\mathrm{E}}}$, where $\bm{\mathcal{S}}_{\mathrm{SIBPN}}^{\bm{\mathcal{Q}}}=\{(\bm{D}_{q,r},M_{q,r})\ |\ r=S,2,\cdots,S_q, q\in\bm{\mathcal{Q}}\}$ and $\bm{\mathcal{Q}}=\bm{\mathcal{Q}}_{\mathrm{T}},\bm{\mathcal{Q}}_{\mathrm{V}},\bm{\mathcal{Q}}_{\mathrm{E}}$.

\subsubsection{The Adoption of 3D Detection Method}
We adopt the \emph{single-stage monocular 3D object detection via keypoint estimation} (SMOKE) in \cite{zliu} as an example to support VBALA, VBALU and VPBCT. To train the SMOKE network, we implement $Q_{\mathrm{D}}$ different vehicle initializations, and let MS take images in $\mathrm{ST}_{1}^{\mathrm{R}}$ and $\mathrm{ST}_{2}^{\mathrm{R}}$ for each vehicle initialization. All the images will be labeled with the 3D bounding boxes of the contained vehicles to construct the samples for the SMOKE network.

\subsection{Performance Metrics}
For VBALA and VBALU, we utilize the achievable transmission rate ratio (ATRR), i.e., the ratio between the average transmission rate achieved by the selected beam pair and the optimal average transmission rate as the performance metric. Denote the $k$th subcarrier channel matrix corresponding to the image set $\bm{\mathcal{I}}_{q,r}$ as $\bm{H}_{q,r,k}$, $r=1,2,\cdots,S_q$, $q=1,2,\cdots,Q$, $k=1,2,\cdots,K$, and denote the shooting moment that MS takes $\bm{\mathcal{I}}_{q,r}$ as $t_{q,r}$. Since $\bm{H}_{q,r,k}$ is assumed unchanged during an interval $T_{\mathrm{d}}$, the optimal transmission rate during the time period $[t_{q,r},t_{q,r}+T_{\mathrm{d}}]$ is given by
\begin{equation}
R(\bm{w}_{\mathrm{B},q,r}^{\mathrm{opt}},\bm{w}_{\mathrm{U},q,r}^{\mathrm{opt}})=\frac{1}{K}\sum_{k=1}^{K}\log_2\left(1+ \frac{P_k}{\sigma^2}|(\bm{w}_{\mathrm{U},q,r}^{\mathrm{opt}})^{\mathrm{H}}\bm{H}_{q,r,k}\bm{w}_{\mathrm{B},q,r}^{\mathrm{opt}}|^2\right),
\end{equation}
where $\bm{w}_{\mathrm{B},q,r}^{\mathrm{opt}}$ and $\bm{w}_{\mathrm{U},q,r}^{\mathrm{opt}}$ are the transmit and receive beam corresponding to the beam pair index $b_{q,r}^{\mathrm{opt}}$ respectively. Then, we assume the communication system will perform beam alignment at moment $t_{q,r}$, $r=1,2,\cdots,S_q$, $q=1,2,\cdots,Q$. Denote the transmit and receive beam selected by VBALA or VBALU at shooting moment $t_{q,r}$ as $\bm{w}_{\mathrm{B},q,r}$ and $\bm{w}_{\mathrm{U},q,r}$ respectively. The ATRR for VBALA or VBALU can be expressed
\begin{equation}
\mathrm{ATRR}_{\mathrm{s}}^{\bm{\mathcal{Q}}}=\frac{\sum_{q\in \bm{\mathcal{Q}}}\sum_{r=1}^{S_q}R(\bm{w}_{\mathrm{B},q,r},\bm{w}_{\mathrm{U},q,r})}{\sum_{q\in \bm{\mathcal{Q}}}\sum_{r=1}^{S_q}R(\bm{w}_{\mathrm{B},q,r}^{\mathrm{opt}},\bm{w}_{\mathrm{U},q,r}^{\mathrm{opt}})},
\end{equation}
where $\bm{\mathcal{Q}}=\bm{\mathcal{Q}}_{\mathrm{V}},\bm{\mathcal{Q}}_{\mathrm{E}}$, and $\mathrm{ATRR}_{\mathrm{s}}^{\bm{\mathcal{Q}}_{\mathrm{V}}}$ and $\mathrm{ATRR}_{\mathrm{s}}^{\bm{\mathcal{Q}}_{\mathrm{E}}}$ represent the ATRR of VBALAA or VBALAU on validation and test set, respectively.

For VPBCT, the BCT prediction accuracy (BCTPA) is used as the performance metric. Denote the BCT predicted from $\bm{D}_{q,r}$ as $M^{'}_{q,r}T_{\mathrm{d}}$. The BCTPA is expressed as
\begin{equation}
\mathrm{BCTPA}^{\bm{\mathcal{Q}}}=\frac{\sum_{q\in \bm{\mathcal{Q}}}\sum_{r=S}^{S_q}\mathbbm{1}^{\mathrm{A}}(M^{'}_{q,r})}{\sum_{q\in \bm{\mathcal{Q}}}(S_q-S+1)},
\end{equation}
where $\mathbbm{1}^{\mathrm{A}}(M^{'}_{q,r})$ is the indicator function
\begin{equation}
  \mathbbm{1}^{\mathrm{A}}(M^{'}_{q,r})=\left\{
  \begin{aligned}
  &1,\ M^{'}_{q,r}=M_{q,r}\\
  &0,\ M^{'}_{q,r}\neq M_{q,r}
  \end{aligned}\right..
\end{equation}
Moreover, there is $\bm{\mathcal{Q}}=\bm{\mathcal{Q}}_{\mathrm{V}},\bm{\mathcal{Q}}_{\mathrm{E}}$, and $\mathrm{BCTPA}^{\bm{\mathcal{Q}}_{\mathrm{V}}}$ and $\mathrm{BCTPA}^{\bm{\mathcal{Q}}_{\mathrm{E}}}$ represent the BCTPA of VPBCT on validation and test set, respectively. We further consider utilizing the ATRR as the performance metric to verify the improvement in transmission rate brought by BCT prediction. We assume the beam pair obtained from beam alignment is always the optimal for simplicity. Similarly, denote the transmit and receive beam adopted at shooting moment $t_{q,r}$ according to the predicted BCTs as $\bm{w}_{\mathrm{B},q,r}^{\mathrm{BCT}}$ and $\bm{w}_{\mathrm{U},q,r}^{\mathrm{BCT}}$ respectively. Since the beam alignment will consume the time $T_{\mathrm{b}}$ in the first interval $T_{\mathrm{d}}$ of each BCT, the ATRR for VPBCT is expressed as
\begin{equation}
\mathrm{ATRR}_{\mathrm{p}}^{\bm{\mathcal{Q}}}=\frac{\sum_{q\in \bm{\mathcal{Q}}}\sum_{r=S}^{S_q}(1-\mathbbm{1}^{\mathrm{R}}(t_{q,r})\frac{T_{\mathrm{b}}}{T_{\mathrm{d}}})R(\bm{w}_{\mathrm{B},q,r}^{\mathrm{BCT}},\bm{w}_{\mathrm{U},q,r}^{\mathrm{BCT}})}{\sum_{q\in \bm{\mathcal{Q}}}\sum_{r=S}^{S_q}R(\bm{w}_{\mathrm{B},q,r}^{\mathrm{opt}},\bm{w}_{\mathrm{U},q,r}^{\mathrm{opt}})},
\end{equation}
where $\mathbbm{1}^{\mathrm{R}}(t_{q,r})$ is the indicator function
\begin{equation}
\mathbbm{1}^{\mathrm{R}}(t_{q,r})=\left\{
\begin{aligned}
&1,\ \mathrm{beam\ alignment\ happens\ during}\ [t_{q,r},t_{q,r}+T_{\mathrm{d}}]\\
&0,\ \mathrm{otherwise}
\end{aligned}\right.,
\end{equation}
$\bm{\mathcal{Q}}=\bm{\mathcal{Q}}_{\mathrm{V}},\bm{\mathcal{Q}}_{\mathrm{E}}$, and $\mathrm{ATRR}_{\mathrm{p}}^{\bm{\mathcal{Q}}_{\mathrm{V}}}$ and $\mathrm{ATRR}_{\mathrm{p}}^{\bm{\mathcal{Q}}_{\mathrm{E}}}$ represent the ATRR of VPBCT on validation and test set, respectively.

\subsection{Simulation Parameters}
\begin{figure}[t]
\centering
\includegraphics[width=0.7\textwidth]{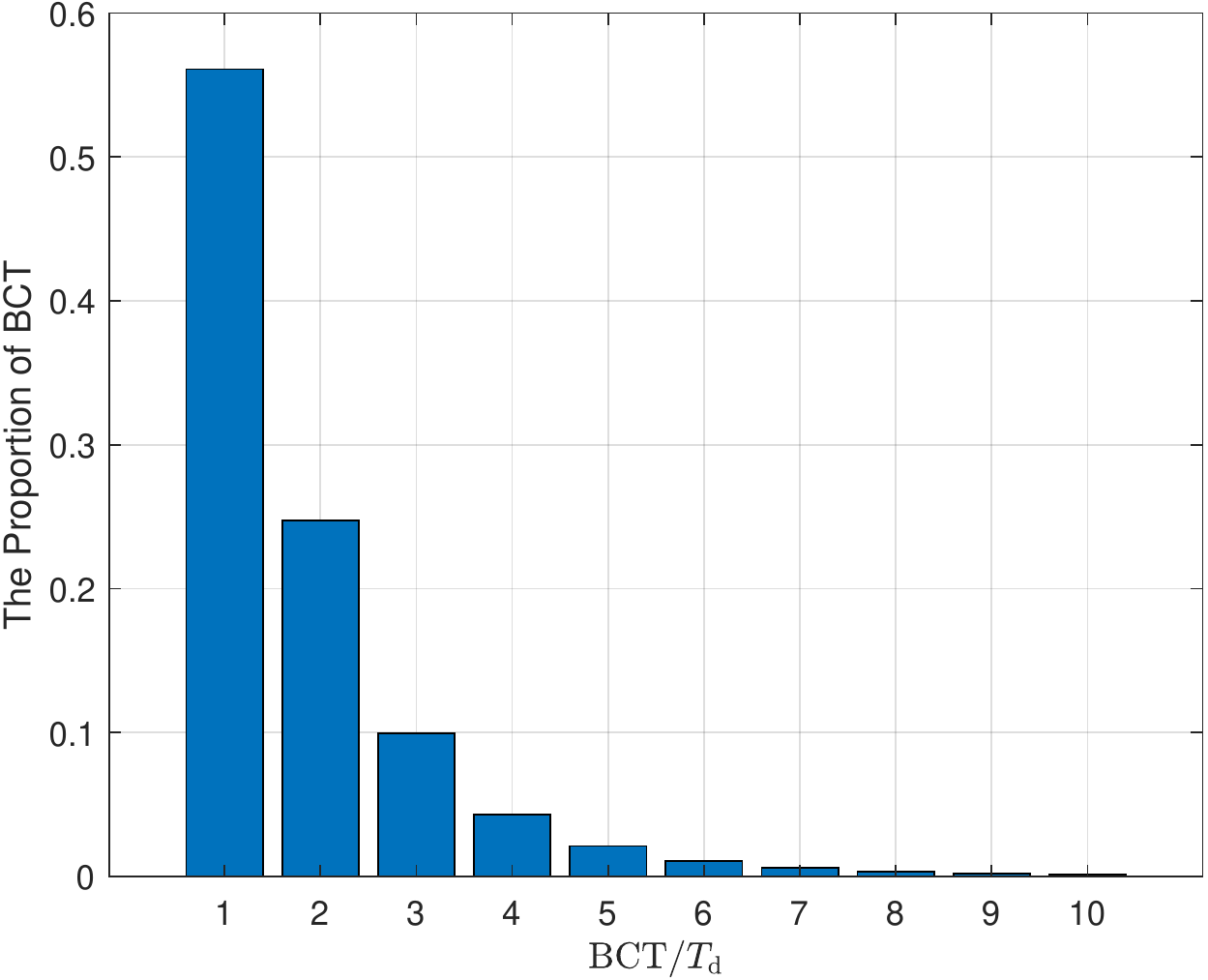}
\caption{The proportions of different BCTs in simulation.}
\end{figure}

\begin{figure}[t]
\centering
\includegraphics[width=0.7\textwidth]{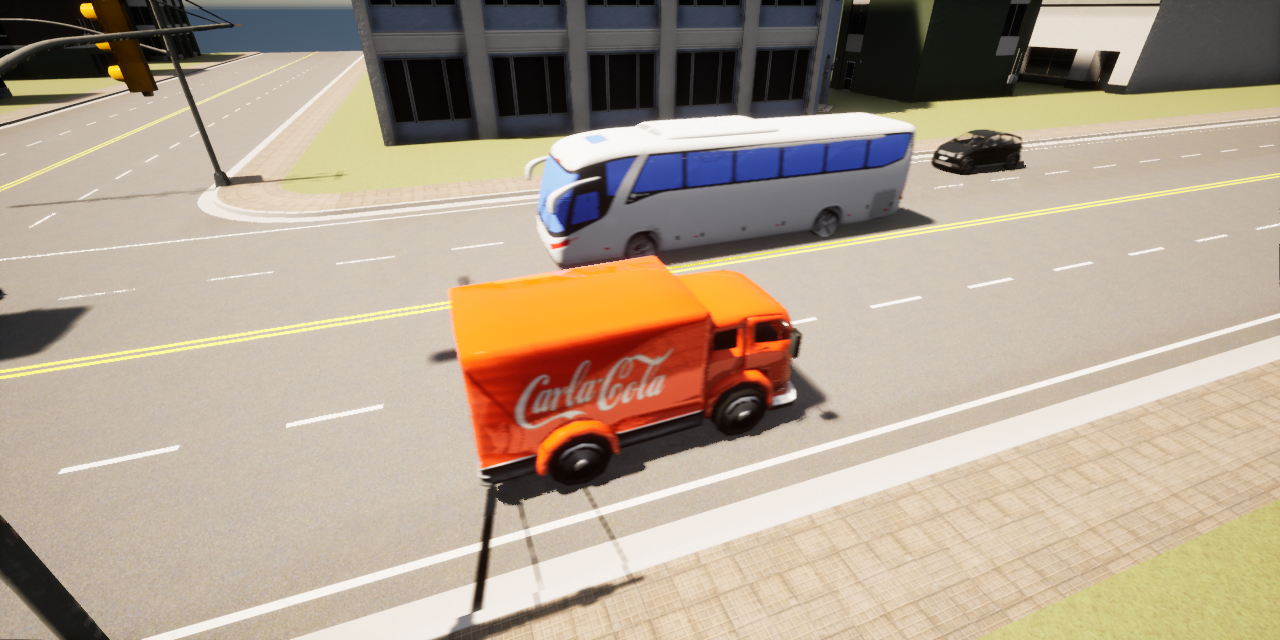}
\caption{A case of the image taken by the camera adopted by BMBA.}
\end{figure}

\begin{table}[t]
\centering
\caption{The Parameters of Layers of SIBPN}
\begin{tabular}{|c|c|c|c|c|}
\hline
Layer Order& Kernel/Pool Size& Strides& Filters &Units\\
\hline
3D Convolution& (1, 7, 7)& (1, 2, 2)& 64& None\\
\hline
3D MaxPooling& (1, 3, 3)& (1, 2, 2)& None& None\\
\hline
3D Convolution& (2, 3, 3)& (1, 1, 1)& 64& None\\
\hline
3D Convolution& (2, 3, 3)& (1, 2, 2)& 128& None\\
\hline
3D Convolution& (2, 3, 3)& (1, 1, 1)& 128& None\\
\hline
Bidirectional ConvLSTM& (3, 3)& (1, 1)& 128& None\\
\hline
2D Convolution& (3, 3)& (2, 2)& 64& None\\
\hline
2D Convolution& (3, 3)& (2, 2)& 32& None\\
\hline
Average& None & None &None& None\\
\hline
FC& None & None &None& 1024\\
\hline
FC& None & None &None& 1024\\
\hline
FC& None & None &None& 3\\
\hline
\end{tabular}
\end{table}

The numbers of antennas at BS and MS are both set as $N_\mathrm{B}=N_{\mathrm{U}}=64$; $K$ is set as 16; $N_\mathrm{B}^{\mathrm{CB}}=N_\mathrm{B}=64;N_{\mathrm{U}}^{\mathrm{CB}}=N_{\mathrm{U}}=64$; $\bm{w}_{\mathrm{B},b}=\bm{a}_{\mathrm{r}}(\frac{2b-2-N_{\mathrm{B}}^{\mathrm{CB}}}{2N_{\mathrm{B}}^{\mathrm{CB}}}\pi)$, $b=1,2,\cdots,N_{\mathrm{B}}^{\mathrm{CB}}$ and $\bm{w}_{\mathrm{U},u}=\bm{a}_{\mathrm{t}}(\frac{2u-2-N_{\mathrm{U}}^{\mathrm{CB}}}{2N_{\mathrm{U}}^{\mathrm{CB}}}\pi)$, $u=1,2,\cdots,N_{\mathrm{U}}^{\mathrm{CB}}$; $P_1=\cdots=P_K$; $\frac{P_k}{K\sigma^2\sum_{q=1}^{Q}S_q}\sum_{q=1}^{Q}\sum_{r=1}^{S_q}\sum_{k=1}^{K}||\bm{H}_{q,r,k}||_{\mathrm{F}}^2=29.5\mathrm{dB}$; $C$ is set to be 4; the camera orientation $\theta^1_{\mathrm{M}}$, $\theta^2_{\mathrm{M}}$, $\theta^3_{\mathrm{M}}$ and $\theta^4_{\mathrm{M}}$ are set to be $0$, $90$, $180$ and $270$ degrees respectively, as shown in Fig.~7; $T_{\mathrm{d}}$ is set as $0.05\mathrm{s}$; $L_{\mathrm{G}}$ and $W_{\mathrm{G}}$ are set as $11.7 \mathrm{m}$ and $2 \mathrm{m}$ respectively; $Q_{\mathrm{D}}$ is set to be 50; $Q$ is set to be 600; $\mathrm{Card}(\bm{\mathcal{Q}}_{\mathrm{T}})$, $\mathrm{Card}(\bm{\mathcal{Q}}_{\mathrm{V}})$ and $\mathrm{Card}(\bm{\mathcal{Q}}_{\mathrm{E}})$ are 480, 60 and 60 respectively; The number of training samples, validation samples and test samples for VDBAN and SIBAN are 27357, 3419 and 3340, respectively.

Due to the limited RSU's coverage area, not all the $N_\mathrm{B}N_{\mathrm{U}}$ beam pairs in $\bm{\mathcal{W}}_{\mathrm{P}}$ can potentially serve as the optimal beam pair. Thus, we count all different optimal beam pairs from all the used beam pairs, i.e., $b_{q,r}^{\mathrm{opt}}$, $r=1,2,\cdots,S_q$, $q=1,2,\cdots,Q$, to form a beam pair set $\bm{\mathcal{W}}_{\mathrm{P}}^{'}$, where $\mathrm{Card}(\bm{\mathcal{W}}_{\mathrm{P}}^{'})=365$. The outputs of VDBAN and SIBAN both are set as the index of the optimal pair in set $\bm{\mathcal{W}}_{\mathrm{P}}^{'}$ instead of $\bm{\mathcal{W}}_{\mathrm{P}}$. Since $\mathrm{Card}(\bm{\mathcal{W}}_{\mathrm{P}}^{'})$ is much smaller than $\mathrm{Card}(\bm{\mathcal{W}}_{\mathrm{P}})$, the simulation overhead and practical network complexity can be reduced.

The number $S$ of image sets used for BCT prediction is set to be 3. The adopted image resolution is $320\times 120$. According to the statistics for $M_{q,r}$, $r=1,2,\cdots,S_q$, $q=1,2,\cdots,Q$, the proportions of different BCTs are obtianed and shown in Fig.~9. It is seen that there exists serious class imbalance problem, since the proportions of short BCTs are much greater than the proportions of long BCTs. Hence, we group the BCTs to reduce the proportion difference between different BCTs. Specifically, we divide all the possible BCTS into three groups $\{1\}$, $\{2\}$, $\{3,4,\cdots,M_{\mathrm{max}}\}$, and set the output of SIBPN as the index of the group containing the accurate BCT. Once a group is predicted by SIBPN, the minimum value of this group will be used as the predicted BCT. Moreover, the re-sampling of the training samples is adopted to further release the class imbalance problem and enhance the generalization performance of SIBPN. Due to the grouping operation, the metric BCTPA is modified as the prediction accuracy of BCT group. When the predicted BCT group from $\bm{D}_{q,r}$ is the group that contain the true BCT $M_{q,r}$, $\mathbbm{1}^{\mathrm{A}}(M^{'}_{q,r})$ is set as 1, otherwise $\mathbbm{1}^{\mathrm{A}}(M^{'}_{q,r})$ is set as 0.

For VDBAN, the VDF $\bm{F}$ is input into a 1-D convolution layer with filter number, kernel size and stride as 1, and is then input into one FC layer with node number $D=64$ to produce $\bm{f}$. The MS's location is also input into one FC layer with node number 64 to produce $\bm{u}$. Then, two transform encoder modules with $d=16$ and $d=32$ are used, and the number of heads of both the two encodes are 4. The final MLP network of VDBAN includes three FC layers with node numbers 1024, 1024 and 365 respectively. The VDBAN is trained for 60 epochs. The network structure of the SIBAN is set as the ResNet101V2 \cite{HeKai}, and the SIBAN is trained for 20 epochs.

For the SIBPN, the parameters of the network structure are shown in the TABLE~V, where the batch normalization and ReLU activation function are used for each convolutional layer. Moreover, through the residual connection, the input of the second 3D convolutional layer is added to the output of this layer, and the input of the third 3D convolutional layer is also added to the output of the fourth 3D convolutional layer. The SIBPN is trained for 20 epochs.
\begin{figure}[t]
\centering
\includegraphics[width=0.7\textwidth]{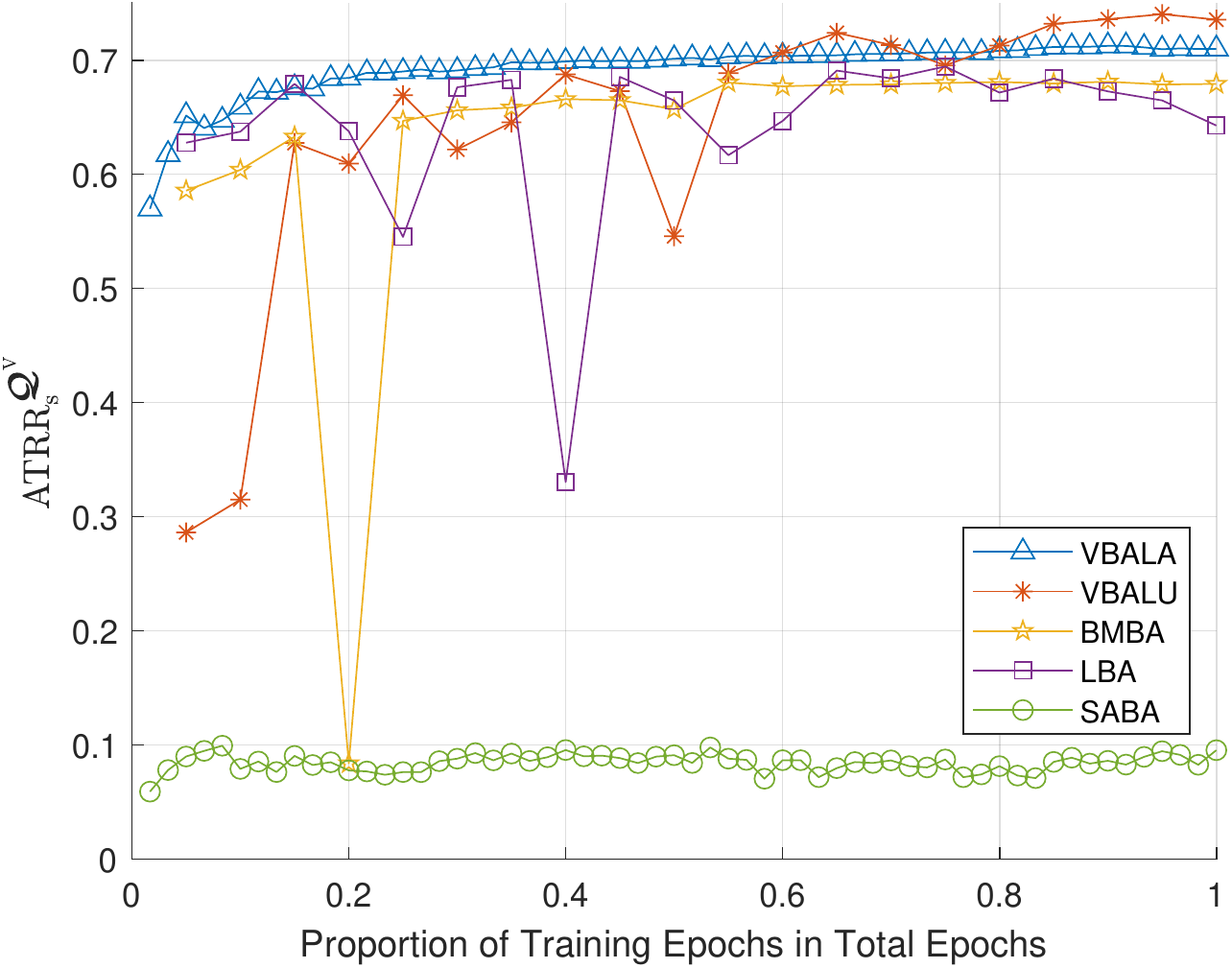}
\caption{${\mathrm{ATRR}}_{\mathrm{s}}^{\bm{\mathcal{Q}}^{\mathrm{V}}}$ achieved by Top-1 beam pair selection with the increase of the number of training epochs.}
\label{VBS_simu1}
\end{figure}

\begin{figure}[t]
	\begin{minipage}[t]{0.5\linewidth}
		\centering
	\includegraphics[width=83mm]{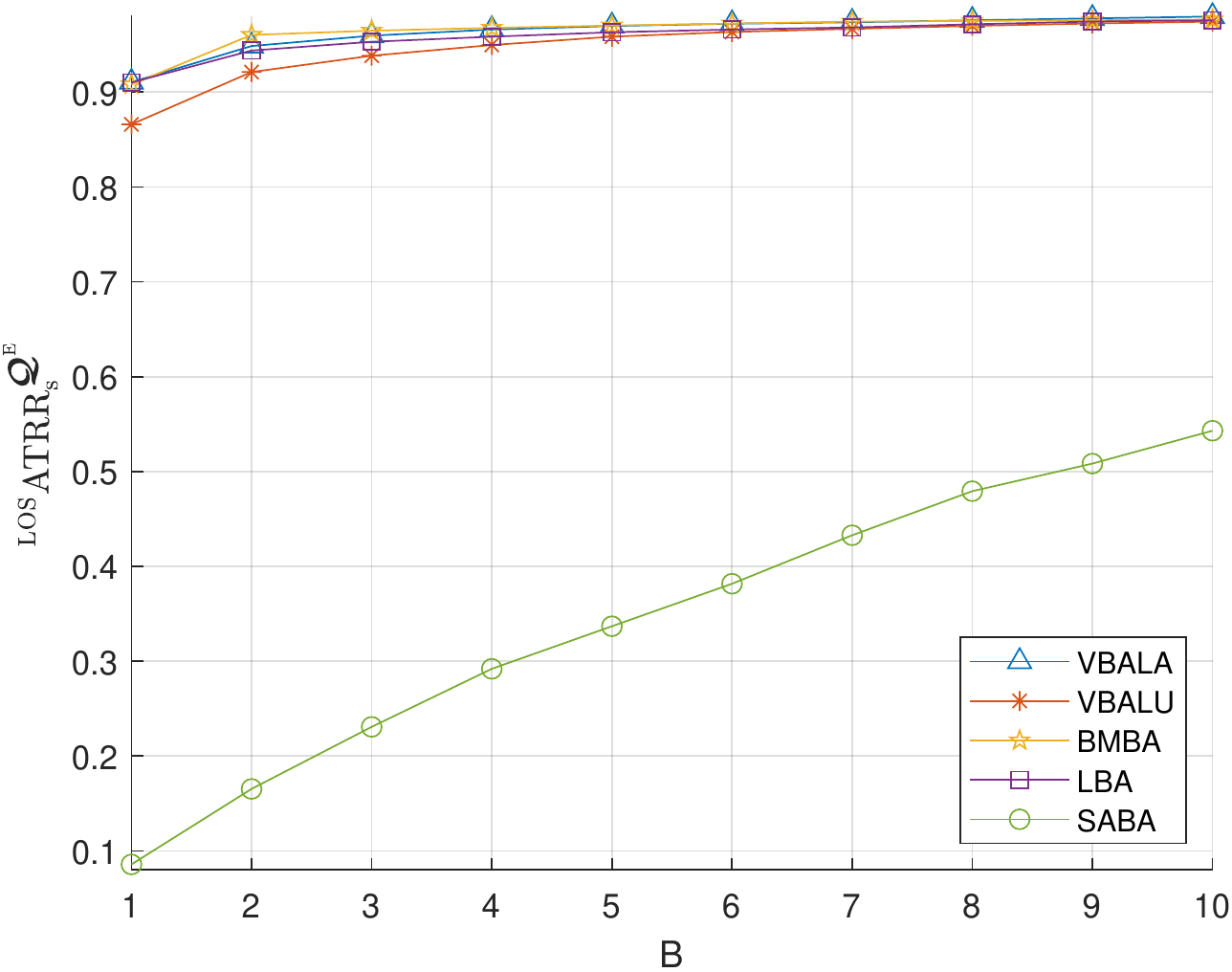}
		\caption{$^{\mathrm{LOS}}\mathrm{ATRR}_{\mathrm{s}}^{\bm{\mathcal{Q}}^{\mathrm{E}}}$ for Top-B beam pair selection. The number of LOS test samples are 1930, which is 58\% of total test samples.}
        \label{VBS_simu2}
	\end{minipage}
	\hspace{1ex}
	\begin{minipage}[t]{0.5\linewidth}
		\centering
			\includegraphics[width=83mm]{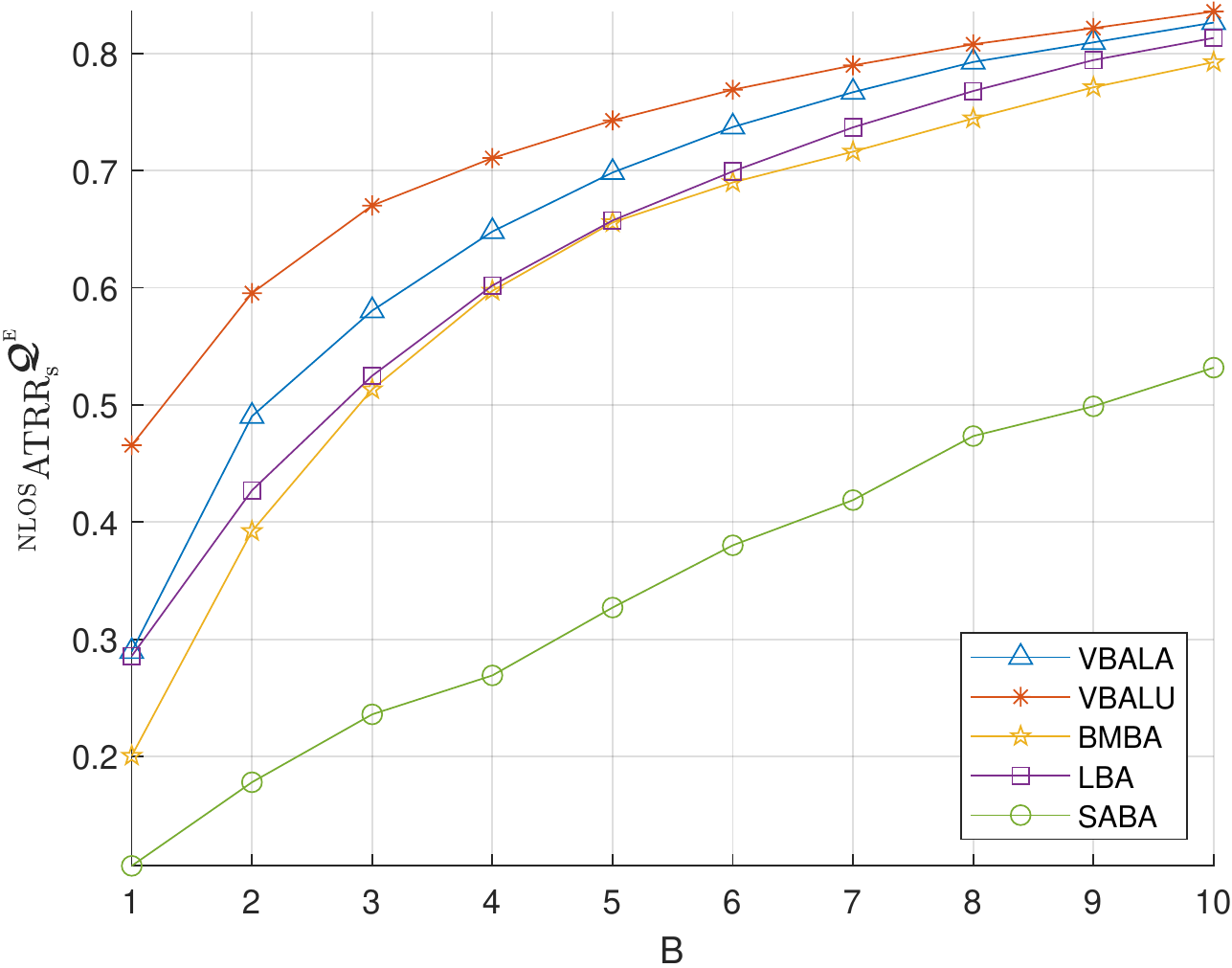}
		\caption{$^{\mathrm{NLOS}}\mathrm{ATRR}_{\mathrm{s}}^{\bm{\mathcal{Q}}^{\mathrm{E}}}$ for Top-B beam pair selection. The number of NLOS test samples are 1410, which is 42\% of total test samples.}
    \label{VBS_simu3}
	\end{minipage}
\end{figure}

\subsection{Results and Discussions}

We compare the VBALA and VBALU, with the BS's vision based multi-modal beam alignment (BMBA) in \cite{GCharan}, the LIDAR based beam alignment (LBA) in \cite{klautau} and the situational awareness based beam alignment (SABA) in \cite{Wang}.

The BMBA uses the images taken at BS and the MS's location as the input features to perform the beam alignment by DNN. For the camera adopted by BMBA, the plane location of the camera is set the same as that of RSU, and the height of the camera is set as $6\mathrm{m}$ to monitor the entire RSU's coverage area. A case of the image taken by the camera adopted by BMBA is shown in Fig.~10. The input features and the network structure is set to be the same as that in \cite{GCharan} to construct the DNN used for BMBA. Moreover, for the DNN of BMBA, the feature extraction subnetwork for images is set to be the same as SIBAN, and the subnetwork for beam alignment after concatenate layer is set to have the same layer sizes and the same number of layers as VDBAN ,i.e., the seven FC layers with node numbers 64, 64, 128, 128, 1024, 1024 and 365 respectively, to keep the fairness. The number of training epochs of the DNN of BMBA is set to be  consistent with SIBAN.

The LBA designs a point cloud feature (PCF) from the MS's location and the point cloud scanned at MS, and utilizes the PCF as the input feature of DNN for beam pair selection. For the DNN of LBA, the input feature is set to be the same as that in \cite{klautau}, while the network structure and the number of training epochs are set to be consistent with SIBAN for fairness. The SABA uses the surrounding vehicles' locations and MS location to present an environmental vehicle location feature (VLF) that is different from the VDF or SIF. Note that \cite{Wang} only uses some classic machine learning methods but not the DNN for beam alignment. For the sake of fairness, we here also adopt the DNN with the same layer sizes, the same number of layers, and the same number of training epochs as VDBAN to infer the optimal beam pair from the VLF. Specifically, the DNN of SABA is constructed by seven FC layers with node numbers 64, 64, 128, 128, 1024, 1024 and 365 respectively. Moreover, the surrounding vehicle locations for VLF are obtained by the 3D detection.

As Fig.~\ref{VBS_simu1} shows, we firstly analyze the ${\mathrm{ATRR}}_{\mathrm{s}}^{\bm{\mathcal{Q}}^{\mathrm{V}}}$ achieved by Top-1 beam pair selection versus the increase of the number of training epochs for the DNNs of VBALA, VBALU, BMBA, LBA and SABA. All the  methods are trained to reach the convergence, and it is seen that the VBALA and VBALU can outperform BMBA, LBA and SABA. Moreover, it is also seen that the ${\mathrm{ATRR}}_{\mathrm{s}}^{\bm{\mathcal{Q}}^{\mathrm{V}}}$ of the SABA is significantly worse than the other four methods and only has a small increase with the increase of the training epochs. This indicates that VLF cannot provide accurate beam alignment compared with the features of the other four methods.

\begin{figure}[t]
\centering
\includegraphics[width=0.7\textwidth]{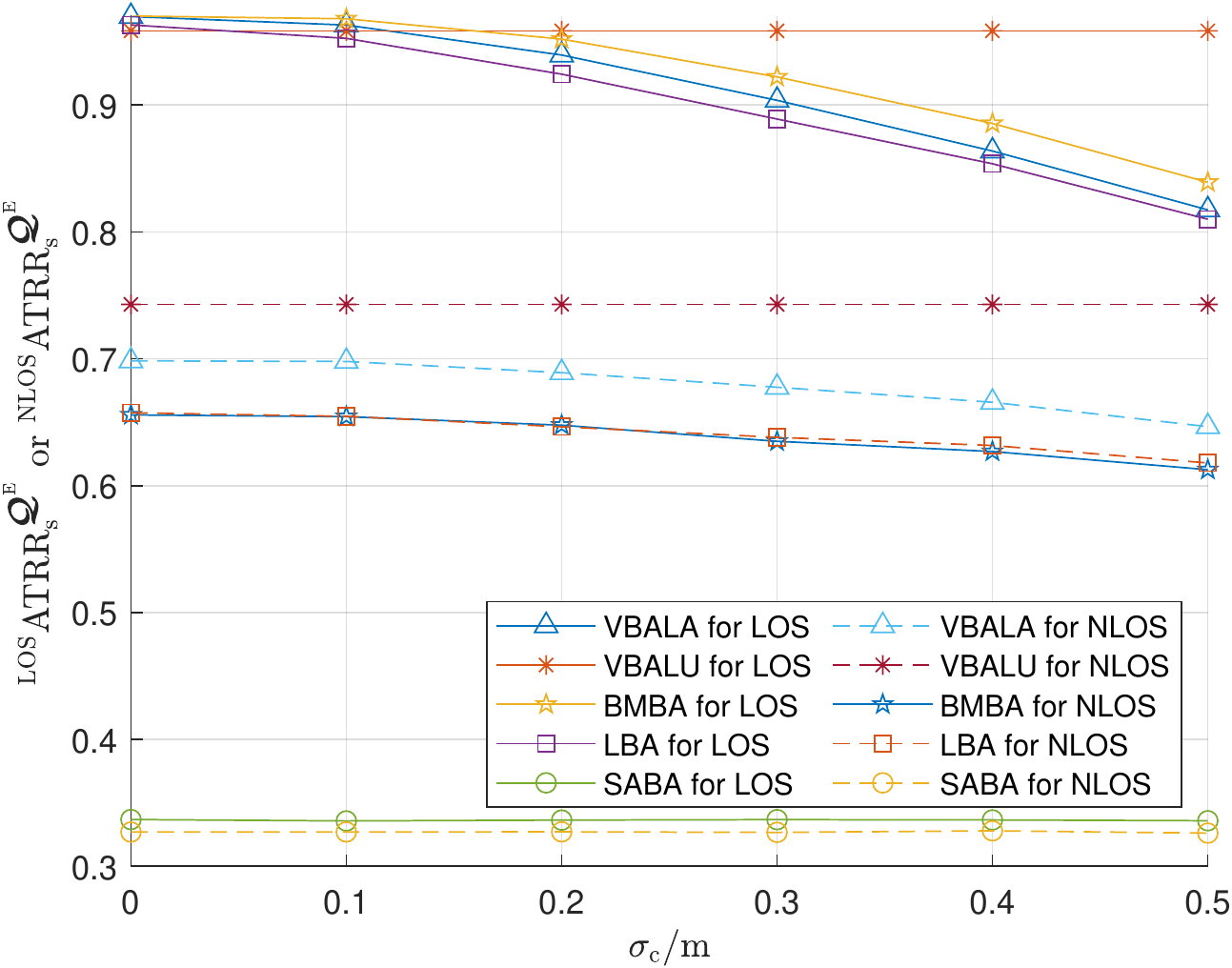}
\caption{$^{\mathrm{LOS}}\mathrm{ATRR}_{\mathrm{s}}^{\bm{\mathcal{Q}}^{\mathrm{E}}}$ and $^{\mathrm{NLOS}}\mathrm{ATRR}_{\mathrm{s}}^{\bm{\mathcal{Q}}^{\mathrm{E}}}$ for Top-5 beam pair selection with different location error $\mathcal{N}(0, \sigma_{\mathrm{c}}^2)$.}
\label{VBS_simu4}
\end{figure}

The optimal weights of the five DNNs are determined by selecting the best epoch according to the ${\mathrm{ATRR}}_{\mathrm{s}}^{\bm{\mathcal{Q}}^{\mathrm{V}}}$ in all epochs. Then, we compare ${\mathrm{ATRR}}_{\mathrm{s}}^{\bm{\mathcal{Q}}^{\mathrm{E}}}$ of VBALA, VBALU, BMBA, LBA and SABA to further confirm their performance differences. Specifically, according to whether the LOS path of channel between RSU and MS is blocked by surrounding objects, we divide all the test samples into two categories: LOS samples and NLOS samples. The number of LOS and NLOS test samples are 1930 and 1410, respectively. Denote the ${\mathrm{ATRR}}_{\mathrm{s}}^{\bm{\mathcal{Q}}^{\mathrm{E}}}$ for LOS and NLOS test samples as $^{\mathrm{LOS}}\mathrm{ATRR}_{\mathrm{s}}^{\bm{\mathcal{Q}}^{\mathrm{E}}}$ and $^{\mathrm{NLOS}}\mathrm{ATRR}_{\mathrm{s}}^{\bm{\mathcal{Q}}^{\mathrm{E}}}$, respectively. The $^{\mathrm{LOS}}\mathrm{ATRR}_{\mathrm{s}}^{\bm{\mathcal{Q}}^{\mathrm{E}}}$ and $^{\mathrm{NLOS}}\mathrm{ATRR}_{\mathrm{s}}^{\bm{\mathcal{Q}}^{\mathrm{E}}}$ achieved by Top-$B$ beam pair selection are shown in Fig.~\ref{VBS_simu2} and Fig.~\ref{VBS_simu3}, respectively. It can be seen that $^{\mathrm{LOS}}\mathrm{ATRR}_{\mathrm{s}}^{\bm{\mathcal{Q}}^{\mathrm{E}}}$ of VBALA is much similar to that of BMBA and LBA, but $^{\mathrm{NLOS}}\mathrm{ATRR}_{\mathrm{s}}^{\bm{\mathcal{Q}}^{\mathrm{E}}}$ of VBALA is better than that of BMBA and LBA. This indicates that though the hardware cost of the camera is lower than the Lidar, VBALA can outperform LBA, and VBALA can also outperform the BMBA without additional feedback overhead of MS location.

The key benefit of VBALA lies in the NLOS scenario, while the simple LOS scenario cannot cause difference between the performance of VBALA, BMBA and LBA. Compared with LBA, the advantage of VBALA under NLOS scenario can be explained as follows: VDF can contain more exact vehicle distribution information, such as the vehicle lengths, widths and heights, whereas LBA simply performs grid quantification for the scanned point cloud to design PDF. Compared with BMBA, the advantage of VBALA is not needed for MS identification. Under the NLOS scenario, since the blockage in the camera view of BS may cause the partial loss of the visual information of MS, the BMBA can not identify and distinguish well the MS and surrounding scatters from the taken image, which leads to the worse performance than the VBALA.

It is also seen that $^{\mathrm{LOS}}\mathrm{ATRR}_{\mathrm{s}}^{\bm{\mathcal{Q}}^{\mathrm{E}}}$ of VBALU is slightly worse than that of VBALA. The reason is that under the LOS scenario, the optimal beam pair depends mostly on the MS's location, but the SIF of VBALU can only reflect the rough MS location by the background information of the images. However, $^{\mathrm{NLOS}}\mathrm{ATRR}_{\mathrm{s}}^{\bm{\mathcal{Q}}^{\mathrm{E}}}$ of VBALU can outperform that of the VBALA, BMBA, LBA and SABA. The reason is that under the NLOS scenario, the distribution information, i.e., the location, size and orientation information, of surrounding scattering objects is more critical than the MS's location information for the beam pair selection, and there is a loss of the location information of surrounding vehicles in the VDF design owing to grid quantization.

Since SIF does not have the grid quantization loss and the VBALU does not need the MS identification, the VBALU can outperform VBALA, BMBA, LBA and SABA for NLOS scenario even without the MS's location. Nevertheless, the computational overhead of SIBAN is higher than VDBAN, since the SIBAN has about $1.22\times 10^{10}$ floating point operations (FLOPs) that is significantly higher than the $3.81\times 10^{6}$ FLOPs of VDBAN. The reason is that SIBAN generally requires larger scale network structure than VDBAN to process SIF whose dimensions are much larger than VDF. Moreover, the generalization of VBALU is worse than that of VBALA, as discussed in Section III.B. In addition, $^{\mathrm{LOS}}\mathrm{ATRR}_{\mathrm{s}}^{\bm{\mathcal{Q}}^{\mathrm{E}}}$ and $^{\mathrm{NLOS}}\mathrm{ATRR}_{\mathrm{s}}^{\bm{\mathcal{Q}}^{\mathrm{E}}}$ of SABA are the worst and are approximately 50\% and 30\% lower than the VBALA and VBALU, respectively, which demonstrates the design of the proposed VDF and SIF is more reasonable than VLF.

We next consider the impact of the MS location estimation error on the $^{\mathrm{LOS}}\mathrm{ATRR}_{\mathrm{s}}^{\bm{\mathcal{Q}}^{\mathrm{E}}}$ and $^{\mathrm{NLOS}}\mathrm{ATRR}_{\mathrm{s}}^{\bm{\mathcal{Q}}^{\mathrm{E}}}$. We compare VBALA, VBALU, BMBA, LBA and SABA under different degrees of location estimation error to evaluate their robustness. The location error is generated as a 2D vector, and the elements of the location error are independent and set to obey the Gaussian distribution $\mathcal{N}(0, \sigma_{\mathrm{c}}^2)$. The $^{\mathrm{LOS}}\mathrm{ATRR}_{\mathrm{s}}^{\bm{\mathcal{Q}}^{\mathrm{E}}}$ and $^{\mathrm{NLOS}}\mathrm{ATRR}_{\mathrm{s}}^{\bm{\mathcal{Q}}^{\mathrm{E}}}$ achieved by Top-5 beam pair selection of all methods under different location error standard deviation $\sigma_{\mathrm{c}}$ are shown in Fig.~\ref{VBS_simu4}. With the increasing of the location error variance, both $^{\mathrm{LOS}}\mathrm{ATRR}_{\mathrm{s}}^{\bm{\mathcal{Q}}^{\mathrm{E}}}$ and $^{\mathrm{NLOS}}\mathrm{ATRR}_{\mathrm{s}}^{\bm{\mathcal{Q}}^{\mathrm{E}}}$ of VBALA are always better than that of LBA. Hence, VBALA has better robustness than LBA under both LOS and NLOS scenarios.

It is seen that the $^{\mathrm{LOS}}\mathrm{ATRR}_{\mathrm{s}}^{\bm{\mathcal{Q}}^{\mathrm{E}}}$ of BMBA is approximately 1.5\% higher on average than the VBALA for different degrees of location estimation error. The reason may be that the images taken at BS will not affected by the location error, whereas the VDF may change due to the location error. However, the $^{\mathrm{NLOS}}\mathrm{ATRR}_{\mathrm{s}}^{\bm{\mathcal{Q}}^{\mathrm{E}}}$ of VBALA can be approximately 4\% higher on average than the BMBA, due to the superiority of MS sensing without the need of MS identification. Thus, the VBALA's overall ATRR for both LOS and NLOS smaples can be higher than BMBA, as indicated in Fig.~\ref{VBS_simu1}. Since the difficulty of beam alignment in the NLOS scenario is significantly higher than that in LOS scenario, as shown in Fig.~\ref{VBS_simu2} and Fig.~\ref{VBS_simu3}, the LOS and NLOS detection can be pre-executed in practice to determine which method to adopt under the scenario with large location error to further improve the ATRR. Moreover, the results reveal the necessity to integrate the visual perception of BS and MS for achieving more robust beam alignment performance.
\begin{figure}[t]
	\begin{minipage}[t]{0.5\linewidth}
		\centering
	\includegraphics[width=83mm]{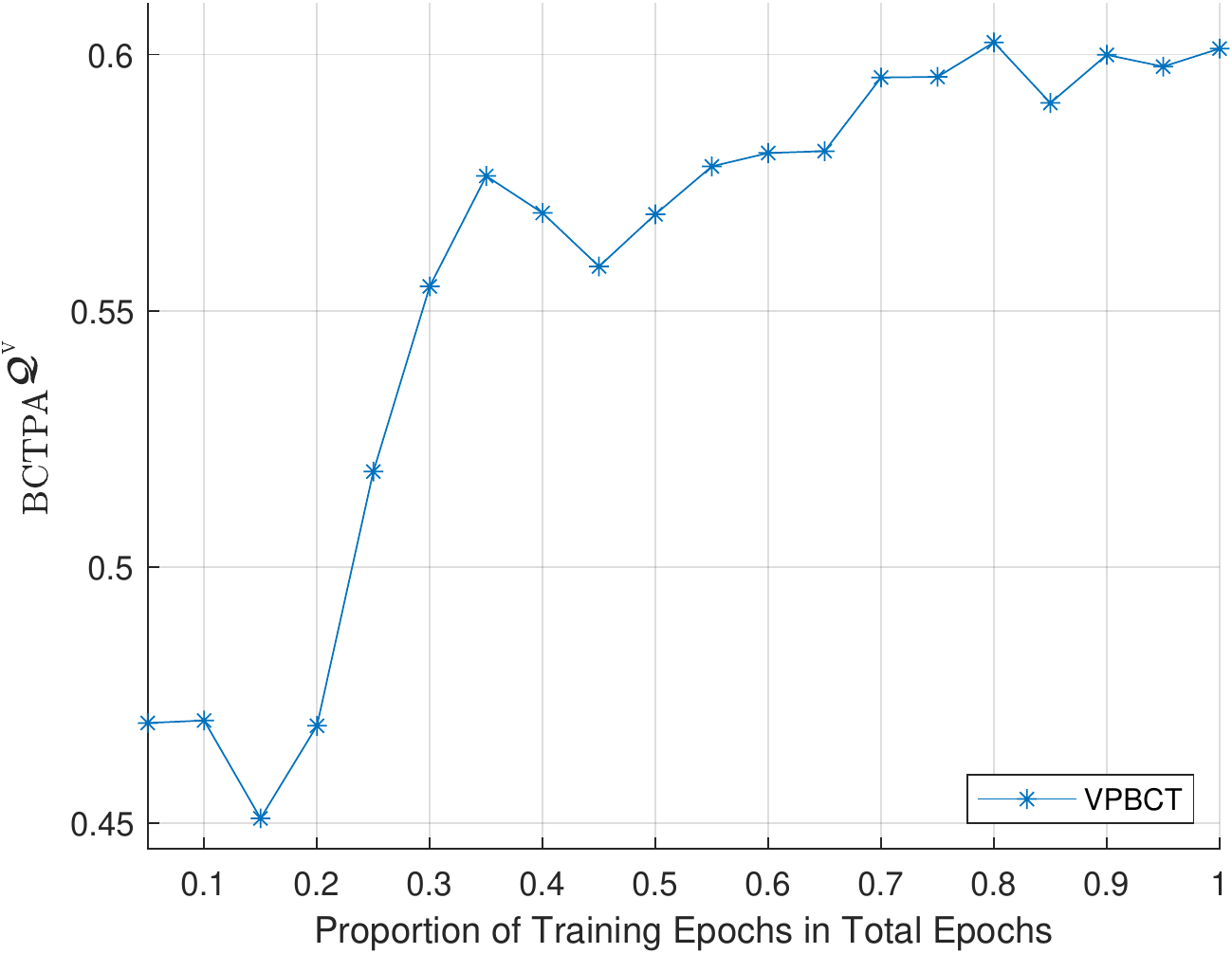}
		\caption{$\mathrm{BCTPA}^{\bm{\mathcal{Q}}_{\mathrm{V}}}$ with the increase of the number of training epochs.}
        \label{VBS_simu5}
	\end{minipage}
	\hspace{1ex}
	\begin{minipage}[t]{0.5\linewidth}
		\centering
			\includegraphics[width=83mm]{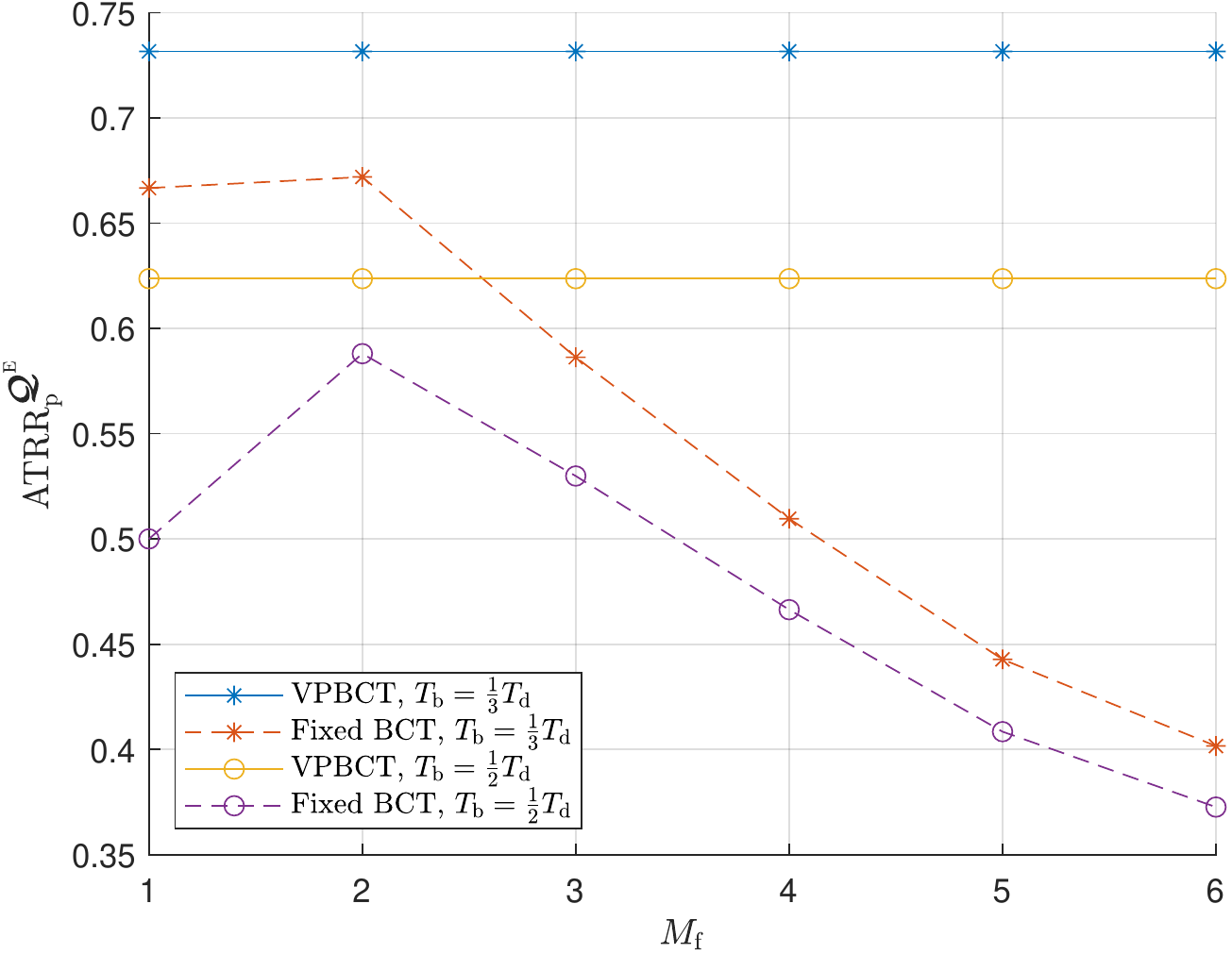}
		\caption{$\mathrm{ATRR}_{\mathrm{p}}^{\bm{\mathcal{Q}}_{\mathrm{E}}}$ under different $M_{\mathrm{f}}$ and $T_{\mathrm{b}}$.}
    \label{VBS_simu6}
	\end{minipage}
\end{figure}

It is interesting that the performance of SABA can remain almost unchanged for LOS and NLOS scenario, even when $\sigma_{\mathrm{c}}$ reaches $0.5\mathrm{m}$. The reason lie in that the SABA mainly adopts $(x^{i,j}_{\mathrm{M}},y^{i,j}_{\mathrm{M}},z^{i,j}_{\mathrm{M}})$, $j=1,2,\cdots,O_i$, $i=1,2,\cdots,C$, to design VLF, and $(x^{i,j}_{\mathrm{M}},y^{i,j}_{\mathrm{M}},z^{i,j}_{\mathrm{M}})$ is obtained from 3D detection and thus is independent of the MS location error. Hence, VLF is almost unaffected by MS location error, which leads to the strong robustness of SABA. However, the performance achieved by SABA is much lower than the other methods. When $\sigma_{\mathrm{c}}>0.16 \mathrm{m}$, both $^{\mathrm{LOS}}\mathrm{ATRR}_{\mathrm{s}}^{\bm{\mathcal{Q}}^{\mathrm{E}}}$ and $^{\mathrm{NLOS}}\mathrm{ATRR}_{\mathrm{s}}^{\bm{\mathcal{Q}}^{\mathrm{E}}}$ of VBALA and BMBA are worse than that of VBALU. This indicates that when the location error is serious, VBALU will perform better than the other four methods.

Next, we analyze the BCT prediction performance of VPBCT. The $\mathrm{BCTPA}^{\bm{\mathcal{Q}}_{\mathrm{V}}}$ versus the increase of the number of training epochs is shown in Fig.~\ref{VBS_simu5}. It is seen that SIBPN is trained to reach convergence and the $\mathrm{BCTPA}^{\bm{\mathcal{Q}}_{\mathrm{V}}}$ can reach about $60\%$. The optimal weight of SIBPN is determined according to the $\mathrm{BCTPA}^{\bm{\mathcal{Q}}_{\mathrm{V}}}$ in all epochs. We compare the $\mathrm{ATRR}_{\mathrm{p}}^{\bm{\mathcal{Q}}_{\mathrm{E}}}$ of VPBCT and the conventional fixed BCT method. The BCT adopted by the fixed BCT method is $M_{\mathrm{f}}T_{\mathrm{d}}$, where $M_{\mathrm{f}}$ is an integer with $M_{\mathrm{f}}\geq 1$. The $\mathrm{ATRR}_{\mathrm{p}}^{\bm{\mathcal{Q}}_{\mathrm{E}}}$ of VPBCT and the fixed BCT method under different $M_{\mathrm{f}}$ and $T_{\mathrm{b}}$ are shown in Fig.~\ref{VBS_simu6}. When $T_b=\frac{1}{3}T_{\mathrm{d}}$ or $T_b=\frac{1}{2}T_{\mathrm{d}}$, the $\mathrm{ATRR}_{\mathrm{p}}^{\bm{\mathcal{Q}}_{\mathrm{E}}}$ of the fixed BCT method is optimal with $M_{\mathrm{f}}=2$, and the optimal $\mathrm{ATRR}_{\mathrm{p}}^{\bm{\mathcal{Q}}_{\mathrm{E}}}$ are $67.2\%$ and $58.8\%$, respectively. However, the $\mathrm{ATRR}_{\mathrm{p}}^{\bm{\mathcal{Q}}_{\mathrm{E}}}$  achieved by the VPBCT are $73.2\%$ and $62.4\%$ respectively for $T_b=\frac{1}{3}T_{\mathrm{d}}$ and $T_b=\frac{1}{2}T_{\mathrm{d}}$. This demonstrates the proposed VPBCT can get higher transmission rate than the fixed BCT method.

\section{Conclusion}
In this paper, we propose two beam alignment methods and a BCT prediction method with the aid of visual perception at MS. The utilization of visual ability from MS's camera can avoid the huge hardware overhead of Lidar and eliminate the privacy concerns and communication cost brought by the visual perception in BS. Under LOS scenario, the performance of VBALA, VBALU, BMBA and LBA is similar. For the NLOS scenario, VBALA and VBALU have the clear advantages in terms of hardware requirements and beam alignment accuracy compared with LBA, and can achieve better beam alignment performance without additional communication cost compared with BMBA. VPBCT can effectively improve the transmission rate by capturing the changes of communication environment from the sequence of scene images. In fact, the visual perception of both BS and MS can only obtain some local observations for the environment. Integrating with the perception ability of different devices with different sensors to enhance the communication performance deserves further research.

\balance

\end{document}